\documentclass{aa}  

\usepackage{graphicx}
\usepackage{txfonts}

\newcommand{\hd}{HD\,96670}
\newcommand{\te}{TESS}
\newcommand{\kms}{km\,s$^{-1}$}

\begin{document}

   \title{Another one (BH+OB pair) bites the dust}
%   \title{Yet another BH+OB pair disappears}

   \author{Ya\"el Naz\'e
          \inst{1}\fnmsep\thanks{F.R.S.-FNRS Senior Research Associate}
          \and
          Gregor Rauw \inst{1}
          }

   \institute{Groupe d'Astrophysique des Hautes Energies, STAR, Universit\'e de Li\`ege, Quartier Agora (B5c, Institut d'Astrophysique et de G\'eophysique), \\
All\'ee du 6 Ao\^ut 19c, B-4000 Sart Tilman, Li\`ege, Belgium\\
              \email{ynaze@uliege.be}
            }

%   \date{Received September 15, 1996; accepted March 16, 1997}

% \abstract{}{}{}{}{}
% 5 {} token are mandatory
 
  \abstract
   {}
   {Most (or possibly all) massive stars reside in multiple systems. From stellar evolution models, numerous systems with an OB star coupled to a black hole would be expected to exist. There have been several claimed detections of such pairs in recent years and this is notably the case of \hd. }
   {Using high-quality photometry and spectroscopy in the optical range, we revisited the \hd\ system. We also examined complementary X-ray observations to provide a broader view of the system properties. }
   {The \te\ light curves of \hd\ clearly show eclipses, ruling out the black hole companion scenario. This does not mean that the system is not of interest. Indeed, the combined analysis of photometric and spectroscopic data indicates that the system most likely consists of a O8.5 giant star paired with a stripped-star companion with a mass of $\sim$4.5\,M$_{\odot}$, a radius of $\sim$1\,R$_{\odot}$, and a surface temperature of $\sim$50\,kK. While several B+sdOB systems have been reported in the literature, this would be the first case of a Galactic system composed of an O star and a faint stripped star. In addition, the system appears brighter and harder than normal OB stars in the X-ray range, albeit less so than for X-ray binaries. The high-energy observations provide hints of phase-locked variations, as typically seen in colliding wind systems. As a post-interaction system, \hd\ actually represents a key case for probing binary evolution, even if it is not ultimately found to host a black hole.  }
   {}


   \maketitle
%
%________________________________________________________________

\section{Introduction}
At the end of their eventful lives, it is commonly believed that the most massive objects of the stellar population ($\sim 20-150$\,M$_{\odot}$)  give rise to stellar-mass black holes (BHs). Following population synthesis models, the Milky Way alone would be expected to harbour thousands, if not millions, of BHs (e.g. \citealt{lam18,ole20}). Furthermore, since most (if not all) massive stars lie in multiple systems, it should not be surprising to find many cases of BHs paired with an OB star or double BH systems. In fact, such detections would prove invaluable for constraining stellar evolution models, especially with respect to binary interactions. However, while double compact systems are now regularly detected during mergers, based on their gravitational wave emissions, identifying BH+OB systems has proven to be elusive despite intense observational efforts.

In binary systems composed of a compact object and a massive star, the former object can accrete material either from the wind of the massive star or following a Roche lobe overflow of that star, leading to copious X-ray emissions. Over a hundred such systems, called high-mass X-ray binaries, are currently known in our Galaxy \citep{for23}. However, amongst them, only Cyg\,X-1 appears to be a secure and undisputed case involving a BH \citep{mil21}. In parallel, since some BH+OB pairs might be X-ray quiet, several recent searches have instead used optical spectroscopy to perform in-depth investigations of massive SB1 systems (e.g. \citealt{mah22}). Several proposals for BH+OB candidates followed, but most detection claims were soon disproved;  for instance, LB-1 \citep{elb20}, HR\,6819 \citep{gie20,bod20}, MWC\,656 \citep{riv24}, NGC\,1850\,BH1 \citep{elb22}, or NGC\,2004\,\#115 \citep{elb22b}. In this paper, we propose to take a new look at such a BH+OB candidate, \hd.

\hd\ is a hot star of the Carina OB2 association, classified by \citet{sot14} as O8.5(n)fp\,var (the `var' suffix added because of varying He\,{\sc ii}\,$\lambda$4686\,\AA). In fact, \hd\ was already noted as variable by \citet{con77} and reported to be a short-period (5\,d) binary by \citet{gar94}, who proposed the first orbital solution. Since only one component was seen in the spectra, only an SB1 solution could be derived. Adding two velocity points from IUE, but discarding low-resolution data from \citet{tha73}, \citet{sti01} refined the orbital solution. Using interferometry, \citet{san14} found a third star in the system, with $\Delta_H\sim1.26$\,mag and $d\sim30$\,mas. Lastly, \citet{gom21} reported photometric variations in addition to the velocity modulation. They interpreted them as ellipsoidal variations (plus a flare). They also found asymmetries in He\,{\sc i} and O lines (photospheric He\,{\sc ii} lines appearing unaffected) as well as a shift between their velocities and those used by \citet{sti01}, both of which they consider as the signature of the third star (the first one being a direct signature, while the second is an indirect signature as it would be the reflex motion due to that third star). They calculated orbital solutions using photometric and spectroscopic variations separately, but they also combined both pieces of information to get a global solution. In both cases, they found that the companion should have a mass of about 6\,M$_{\odot}$, although no trace of it was detected in the optical spectra. This absence of optical signature, combined with the detection of hard X-rays with NuSTAR, led them to conclude that \hd\ most probably contains a BH companion.

To revisit the properties of \hd, we use optical and X-ray data, first presented in Sect. 2 and then analysed in Sect. 3. These observational constraints lead to a re-assessment of the system's nature, which is discussed in Sect. 4.

\section{Data}

\subsection{Stellar properties}
In the Gaia-DR3 catalog \citep{bai21}, \hd\ displays a distance of $3282_{-288}^{+312}$\,pc. The Gaia parallax appears secure (small error as $R_{plx}=\pi/\sigma_\pi=11.1$ and good quality as $RUWE=1.2$). This is in line with other determinations \citep{gar94}, but \citet{gom21} nevertheless preferred to use the distance from a set of neighbouring Carina stars (2.83\,kpc). In Simbad, \hd\ appears with magnitudes $B=7.60$ and $V=7.43$. Its reddening has been evaluated to $E(B-V)=0.39$\,mag by \citet{bow08} and 0.43\,mag by \citet{sav01} or from the $B-V$ and the calibration of \citet{mar06} for O8.5 stars. This leads to absolute magnitudes of $M_V=-6.04$ for the lowest extinction and distance and $M_V=-6.48$ for the largest ones. \citet{mar05} listed $M_V$ values of --4.19, --5.32, and --6.29 for O8.5 V, III, I stars, respectively. On this basis alone, \hd\ would thus be classified as a supergiant and this luminosity class sometimes appears in the literature \citep{sav01,san14}. Using the bolometric correction for supergiants, we would then derive a bolometric luminosity of $\log(L_{\rm BOL}/L_\odot)=5.47-5.65$ for \hd ; it would be 5.51--5.69 when adopting the bolometric correction for giants. 

\subsection{\te}

\hd\ was  observed over five months by the \te\ satellite \citep{ric15}, in Sectors 10, 11, 37, 63, and 64. Sky images were taken with a cadence of 30\,min in the first two Sectors, 10\,min for Sectors 37, and 200\,s for the last two Sectors. The sector duration leads to a natural peak width in frequency space of 1/25\,d; namely, 0.04\,d$^{-1}$. The \te\ data were reduced and calibrated using a pipeline inspired by that of the {\it Kepler} mission. Individual light curves were then derived from image cutouts of 51$\times$51 pixels using aperture photometry performed with the Python package Lightkurve\footnote{https://docs.lightkurve.org/}. The source mask was defined from pixels with fluxes above 30 times the Median Absolute Deviation over the median flux, while the background mask was defined by pixels with fluxes below the median flux (i.e. below the null threshold). To estimate the background, two methods were used: (1) a principal component analysis (PCA), with five components; (2) a simple median calculation. In each case, the two background-subtracted light curves were compared. The PCA-corrected light curves were similar or better (less long-term trends) and they were thus retained. The \te\ fluxes were converted into magnitudes and the average magnitude was subtracted from each individual light curve.

Since \te\ has a rather coarse PSF, we examined the environment of \hd\ in Gaia-DR3. Within a one arcmin radius (about 3 \te\ pixels, typical of the chosen apertures), all other objects display magnitude differences of $\Delta(G)>4$\,mag. The ten brightest neighbours have $\Delta(G)=4-8$\,mag, with separations of 0.5--1\arcmin. Thus, only scarce contamination is expected from these stars.

\subsection{FEROS}
The high-resolution echelle spectrograph FEROS, installed on the 2.2\,m telescope at La Silla observatory (European Southern Observatory, ESO), was used to observe \hd\ nine times. Eight spectra, taken between 2004 and 2009 (Obs ID 074.D-0300, 076C-0431, 076D-0294, and 083.D-0589), are publicly available and were downloaded from the ESO archives\footnote{http://archive.eso.org/cms.html}. Their signal-to-noise ratios (S/Ns) lie between 170 and 350. We note that the heliocentric correction was directly applied to these spectra. The spectra were normalized in several regions harbouring the main photospheric lines (4000--4600, 4600--5100, 5350--5950, and 6400-6700\,\AA). 

\subsection {X-ray data}
The {\it Chandra} X-ray telescope observed \hd\ slightly off-axis in October 2020 (ObsID 23643). Its spectrum was built using CIAO v4.16. The X-ray data were extracted around the Simbad coordinates of \hd\ in a circle of 10 pixels radius and the background was taken in the surrounding annulus with an outer radius of 30 pixels. A minimum number of counts of 10 was used to bin the spectrum and weighted spectral matrices were built during data extraction using the {\sc specextract} task.

The {\it Swift} X-ray telescope observed \hd\ in PC mode nine times. The light curve was built using the on-line tool\footnote{https://www.swift.ac.uk/user\_objects/} in the 0.5--10.\,keV band. The number of collected counts is low so no individual spectroscopic analysis can be performed and only a combined spectrum was built using the same tool, then the same grouping as for {\it Chandra} was applied. These {\it Swift} data must be taken with caution. Indeed, the X-ray measurements suffer from optical contamination if the targets are optically bright: the limit for hot stars is around $V=8.4,$ while \hd\ has $V=7.4$. One may note that the photometric changes (of a few mmag) are too small to significantly affect this contamination. This being said, we stress that the combined {\it Swift} spectrum shows no strong excess at very low energies, as is typically recorded in case of optical loading. Therefore, we decided to (cautiously) examine the {\it Swift} data. Extraction was however performed considering all grades but also keeping only $grade=0$, since this yields more reliable data and guarantees the absence of optical contamination (although it lowers the S/N). 

\begin{figure*}
  \centering
  \includegraphics[width=6cm]{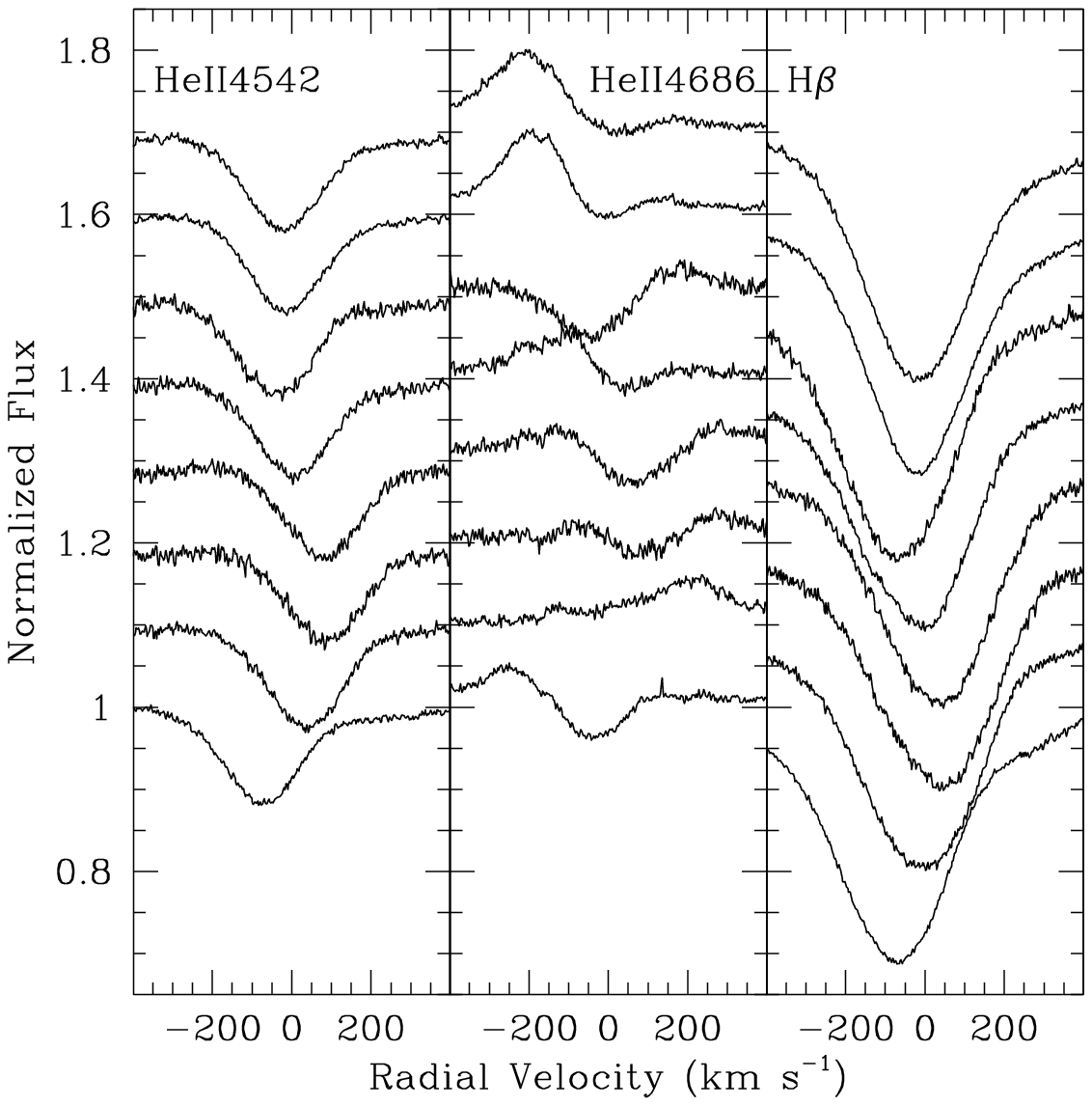}
  \includegraphics[width=6cm]{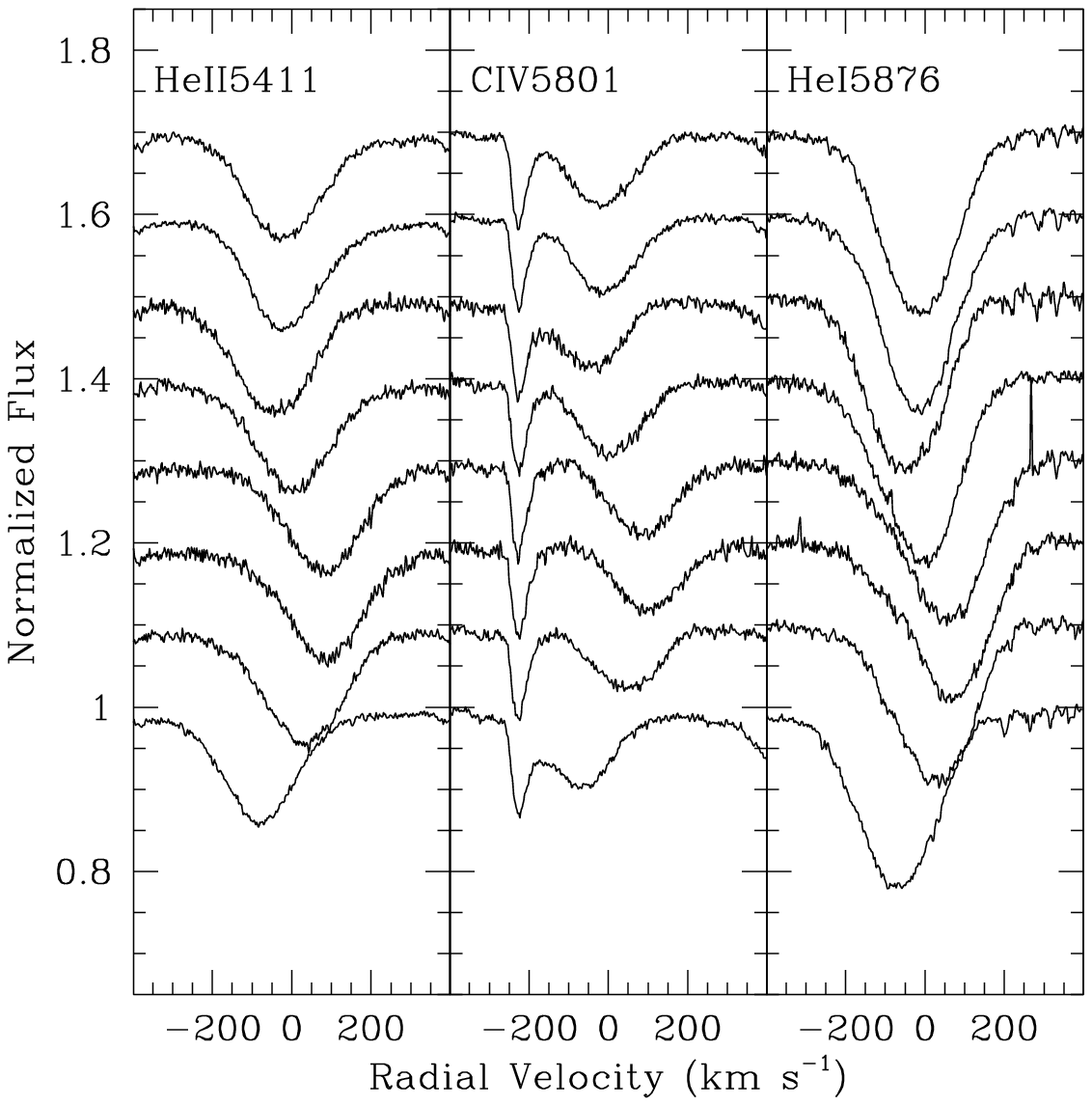}
  \includegraphics[width=6cm]{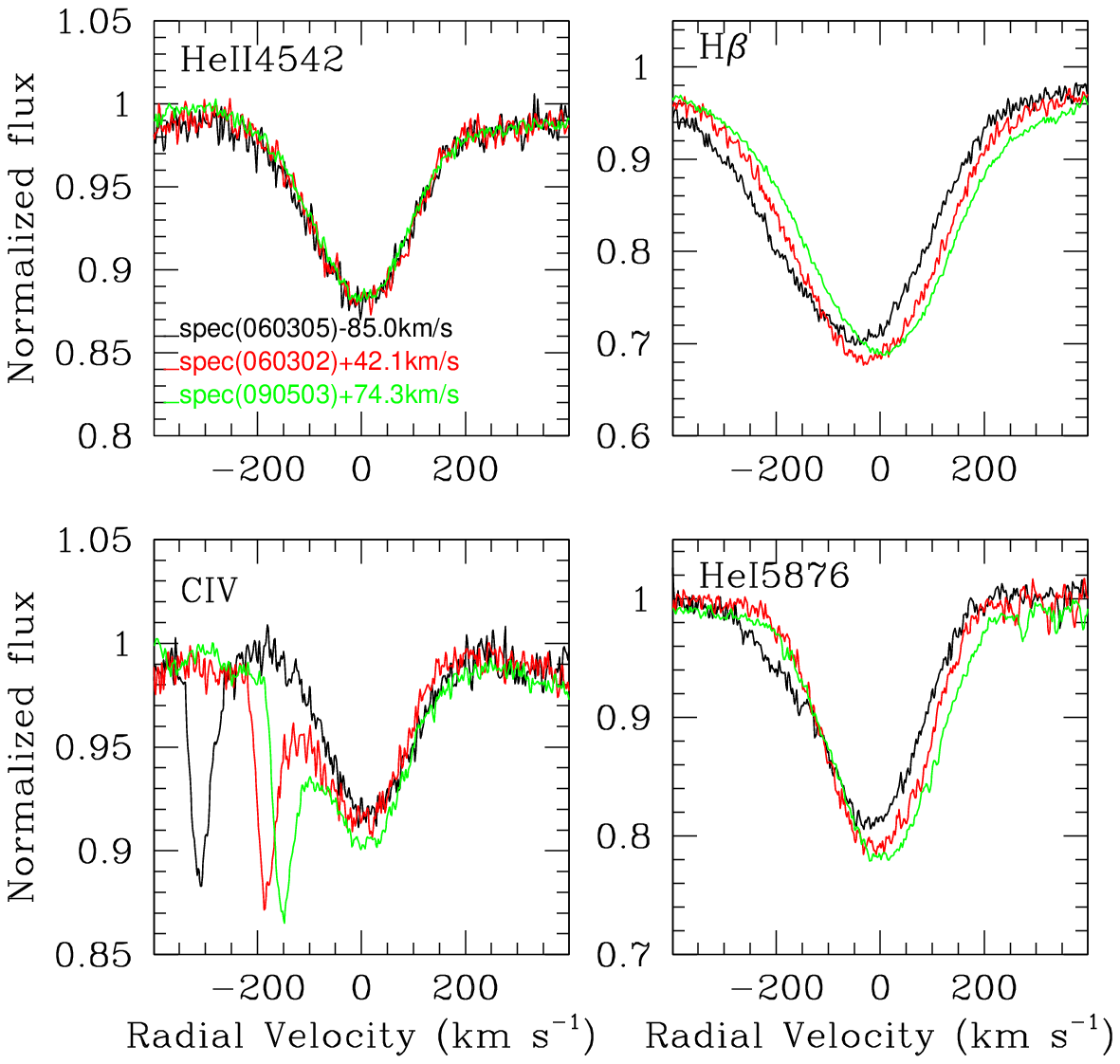}
  \caption{{\it Left and middle panels:} Various lines recorded in the eight FEROS spectra, with arbitrary vertical shifts to make differences clearer (oldest spectrum on top, earliest at bottom). {\it Right:} Lines corrected for the average He\,{\sc ii} velocity (Table \ref{rv}):  He\,{\sc ii} and C\,{\sc iv} display no profile change, while H and He\,{\sc i} do. 
  }
  \label{raies}
\end{figure*}

\section{The \hd\ system}
\subsection{Constraints from velocities}
In the FEROS spectra (Fig. \ref{raies}), the profiles of some spectral lines show no morphological changes but display changing Doppler shifts, such as He\,{\sc ii} absorption lines, which were known to be unaffected by contamination \citep{gom21}. In contrast, these authors found Balmer and He\,{\sc i} lines to be contaminated by the third star and we confirm here their asymmetric appearance, at some phases. Finally, some lines display a complex behaviour. This is notably the case of H$\alpha$ or He\,{\sc ii}\,$\lambda$4686\AA\ (Fig. \ref{raies}). The profile of these lines is a combination of emission and absorption. In the spectra of \citet[see their Figure 8]{gom21}, the He\,{\sc ii} emission always appeared at $\sim200$\,\kms, on the red-shifted side. These authors interpreted it as `a complex accretion wind'. In some of our profiles as well as in those shown by \citet[see their Figure 7]{sot14}, emission appears on both the red-shifted and the blue-shifted sides. Thus, the line profile is far from constant, with complex variations. 

\begin{figure}
  \centering
  \includegraphics[width=9cm]{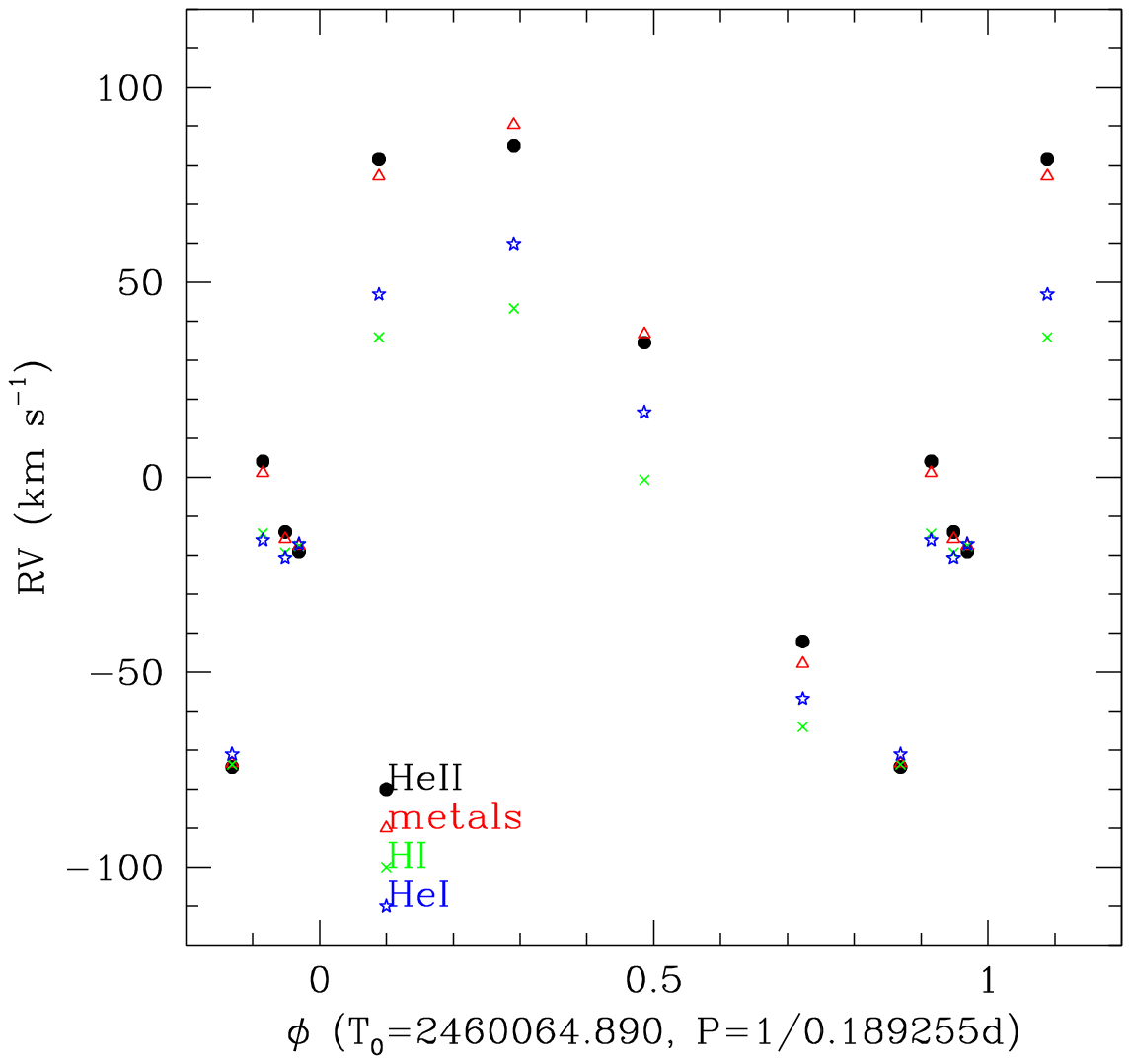}
  \caption{RVs measured on FEROS data from different sets of lines.
  }
  \label{rvlines}
\end{figure}

The radial velocities (RVs) were measured by fitting Gaussians to the bottom part of the photospheric He\,{\sc ii} lines at 4199.83, 4541.59, and 5411.53\AA. For completeness, we also measured the velocities of two Balmer lines in the same way  (H$\gamma$\,$\lambda$4340.468\AA\ and H$\beta$\,$\lambda$4861.33\AA), three He\,{\sc i} lines (at 4026.072, 4471.512, 5875.62\AA), and several metallic lines (Si\,{\sc iv}\,$\lambda$4088.863\AA, N\,{\sc iii}\,$\lambda$4379.09,4634.25\AA, O\,{\sc iii}\,$\lambda$5592.252\AA, and C\,{\sc iv}\,$\lambda$5801.34,5811.97\AA, which are all  in absorption, except N\,{\sc iii}\,$\lambda$4634). The velocities from photospheric lines of He\,{\sc ii} and metals are in excellent agreement, but differences are sometimes detected with the other lines (Fig. \ref{rvlines}). Table \ref{rv} provides average velocities from lines of H, He\,{\sc i}, He\,{\sc ii}, and metals (only absorptions). As they appear the most secure, RVs from photospheric He\,{\sc ii} lines will be considered as our reference velocities, as done by \citet{gom21}. A period search algorithm \citep{HMM,gos01} was applied to the full RVs set, namely, by complementing data from \citet{con77}, \citet{gar93}, \citet{sti01}, and \citet{gom21} with our own. The best frequency was found at $0.189255\pm0.000005$\,d$^{-1}$, corresponding to a period of $5.28388\pm0.00018$d in line with that of Gomez \& Grindlay ($P=5.28388\pm0.00046$\,d).

\begin{figure*}
  \centering
  \includegraphics[width=9cm]{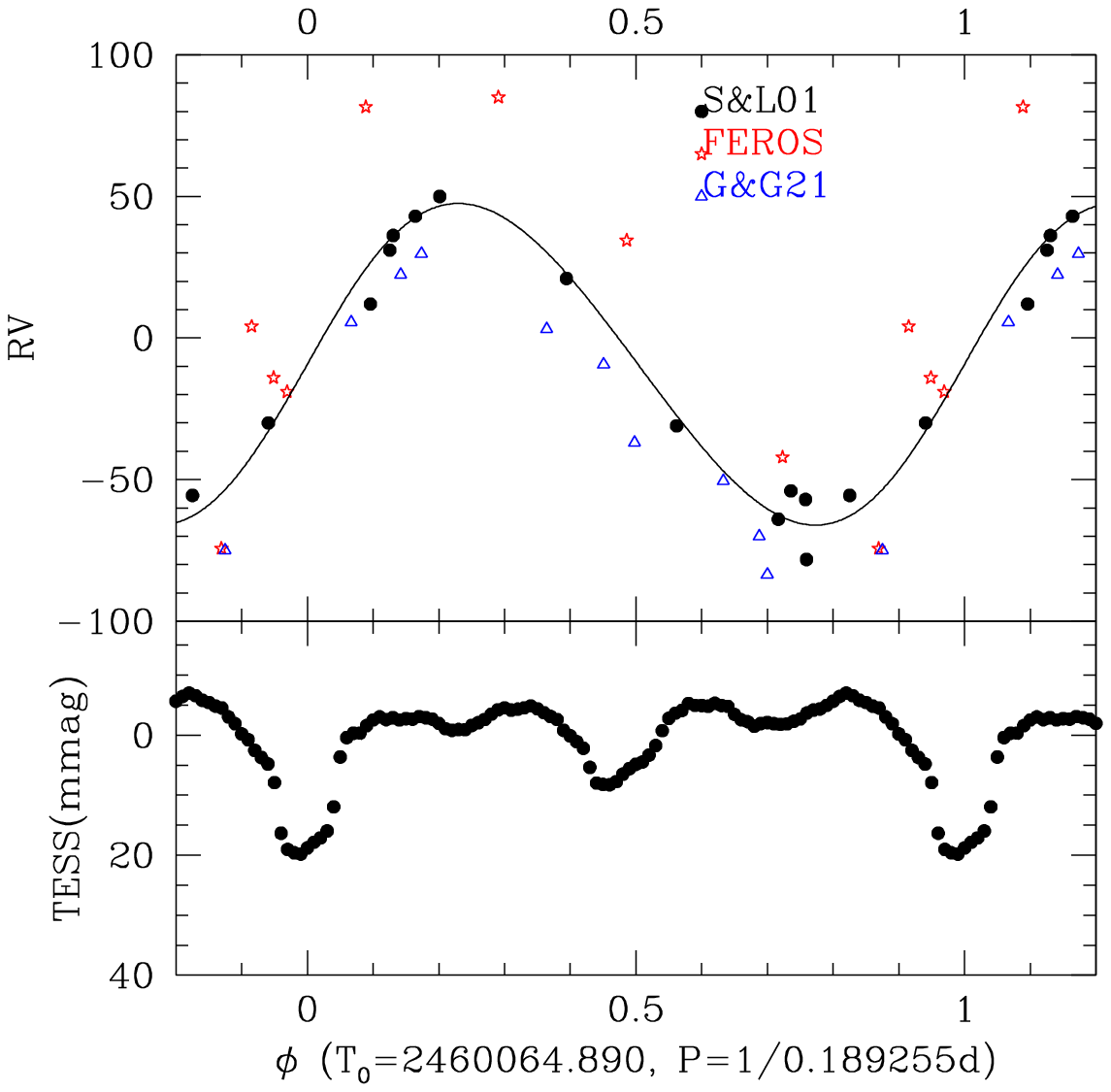}
  \includegraphics[width=9cm]{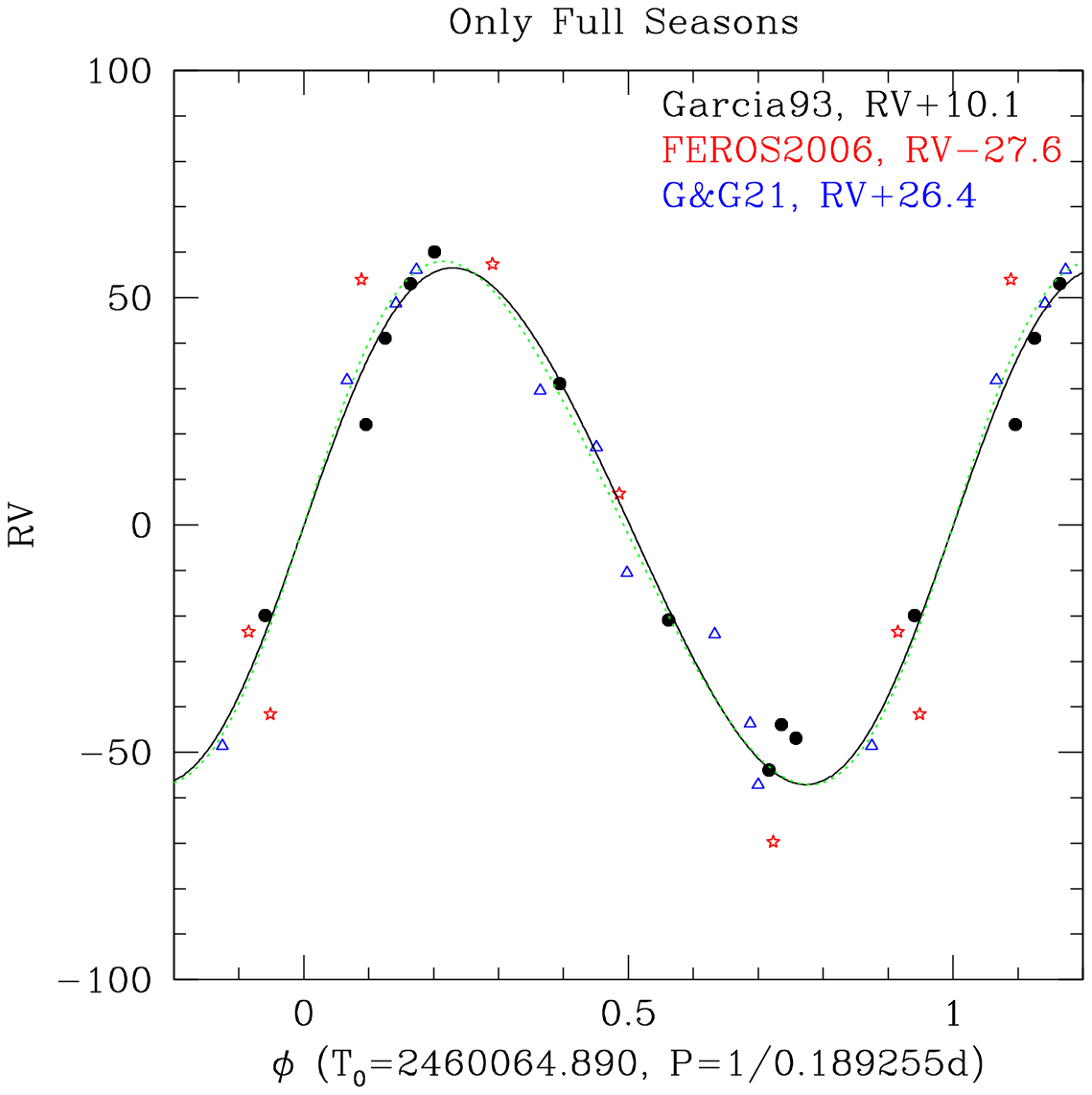}
  \caption{{\it Left:} RVs of \hd\ used by Stickland \& Lloyd (2001:\ black dots), averages of the RVs derived by Gomez \& Grindlay (2021:\ blue triangles), and RVs measured on FEROS data (red stars). The Stickland \& Lloyd solution is also shown - beware that the period used by these authors was different. The bottom panel displays the average \te\ LC, phased with the same ephemeris. {\it Right:} Shifted RVs along with the solutions of Stickland \& Lloyd (black line, with $\gamma$ set to zero) and ours (dotted green line). Here, $\phi=0$ corresponds to conjunction with the O star in front and only the FEROS points from 2006 (marked with an asterisk in Table \ref{trend}) are shown.
  }
  \label{rvphot}
\end{figure*}

\begin{figure}
  \centering
  \includegraphics[width=9cm]{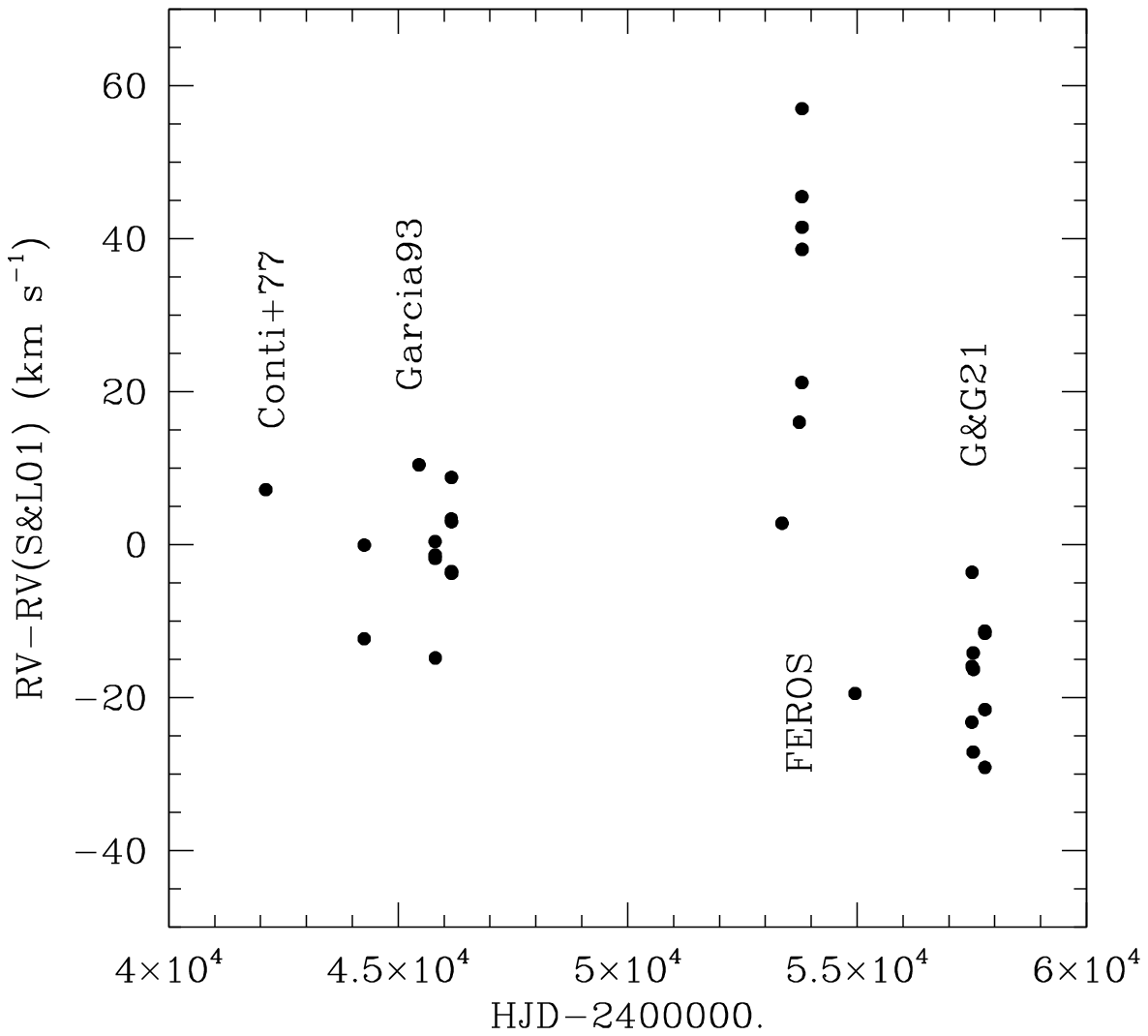}
  \caption{Difference between RVs of \hd\ and the solution of Stickland \& Lloyd (2001).
  }
  \label{trend}
\end{figure}

\begin{table}
  \scriptsize
  \caption{RVs of \hd.
 \label{rv}}
  \begin{tabular}{lll}
    \hline
HJD-2\,400\,000. & RV (\kms) & Ref \\
\hline
42114.64   & $-55.6\pm3.0$ & Conti et al. 1977 \\
44259.549  & -78.2 & Stickland \& Lloyd 2001 \\
44261.509  &  36.2 & Stickland \& Lloyd 2001 \\
45453.576  & $-54\pm3$   & Garcia 1993 \\
45804.577* &  $43\pm4$   & Garcia 1993 \\
45805.794* &  $21\pm4$   & Garcia 1993 \\
45807.498* & $-64\pm6$   & Garcia 1993 \\
45809.500* &  $12\pm6$   & Garcia 1993 \\
46158.793* &  $50\pm5$   & Garcia 1993 \\
46160.698* & $-31\pm4$   & Garcia 1993 \\
46161.735* & $-57\pm5$   & Garcia 1993 \\
46162.698* & $-30\pm3$   & Garcia 1993 \\
46163.673* &  $31\pm8$   & Garcia 1993 \\
53364.774  & $-17.6/-17.1/-19.0\pm3.1/-17.6$ & this work \\
53739.817* & $-19.3/-20.6/-14.0\pm4.7/-15.8$ & this work \\
53796.751* & $-64.0/-56.8/-42.1\pm3.1/-47.8$ & this work \\
53797.769* & $-14.4/-16.1/  4.1\pm1.4/  1.1$ & this work \\
53798.685* &  $35.9/ 46.9/ 81.6\pm1.3/ 77.3$ & this work \\
53799.750* &  $43.3/ 59.8/ 85.0\pm0.6/ 90.2$ & this work \\
53800.783* &  $-0.6/ 16.7/ 34.5\pm1.8/ 36.7$ & this work \\
54954.691  & $-73.6/-71.0/-74.3\pm4.0/-73.4$ & this work \\
57500.624* & $-83.6\pm1.2$ & Gomez \& Grindlay 2021 \\
57504.594* &  $-9.3\pm2.3$ & Gomez \& Grindlay 2021 \\
57505.553* & $-50.4\pm1.0$ & Gomez \& Grindlay 2021 \\
57529.547* &  $29.7\pm1.5$ & Gomez \& Grindlay 2021 \\
57530.555* &   $3.1\pm1.2$ & Gomez \& Grindlay 2021 \\
57534.665* &  $22.3\pm1.8$ & Gomez \& Grindlay 2021 \\
57782.606* &   $5.5\pm2.1$ & Gomez \& Grindlay 2021 \\
57784.889* & $-36.9\pm4.6$ & Gomez \& Grindlay 2021 \\
57785.891* & $-70.1\pm1.2$ & Gomez \& Grindlay 2021 \\
57786.881* & $-75.0\pm1.7$ & Gomez \& Grindlay 2021 \\
\hline
  \end{tabular}
\tablefoot{As the Gomez \& Grindlay original RV points (available at CDS) were often taken within $<$0.1\,d, or $<$0.02 in phase, we performed daily averages of these data, which resulted in ten independent RVs - their errors correspond to the 1-$\sigma$ dispersion. Our RVs correspond to averages on Balmer lines (H$\gamma$ and H$\beta$) / He\,{\sc i} lines (at 4026, 4471, 5876\AA); He\,{\sc ii} lines (at 4200, 4542, 5412\AA); and metallic lines (Si\,{\sc iv}$\lambda$4088, N\,{\sc iii}$\lambda$4379, O\,{\sc iii}$\lambda$5592, C\,{\sc iv}$\lambda$5801,5812). Quoted errors correspond to the 1-$\sigma$ dispersion of the three He\,{\sc ii} measurements. A * indicates data taken within the same year and considered for the determination of an orbital solution, after shifting (see text for details).  }
\end{table}

In each study, different sets of lines are used, so that shifts between them may occur. In addition, the presence of a third body could also lead to velocity shifts from epoch to epoch. \citet{gom21} reported a shift of 16.7\,\kms\ between their data (spread over a year) and those of previous studies (spread over 4000\,d). Our FEROS velocities (spread over 1600\,d) also display such shifts, as can be seen in the left panel of Figure \ref{rvphot}. In this figure, phase zero corresponds to the deeper eclipse (see Sect. 3.2), namely, to a conjunction. The orbital solution of \citet{sti01}, also shown on the figure, has a $T_0$ corresponding to periastron and, from the orbital solution, we can calculate that the conjunction occurs at $\phi_c=0.097$. Thus, we  shifted the Stickland \& Lloyd solution RV curve by -0.097 to display it on Fig. \ref{rvphot}.

We further used the Stickland \& Lloyd solution to calculate the expected RVs for the phase of each spectroscopic observation, calculated from our period and $T_0$, plus considering the above phase shift. The observed minus predicted RV residuals do not show any coherent trend with time, as would be expected for reflex motion due to the presence of a third body (Fig. \ref{trend}). However, we should not exclude the possibility that the noise is hiding small orbital variations. Clearly, a long-term, high quality monitoring of \hd\ is needed to assess whether reflex motion is present and (if it is) to constrain the orbit around the third body. 

We then considered observing sets taken within a year (see * in Table \ref{rv}); namely, we discarded isolated data points and calculated the average differences in each yearly set. We found average shifts $RV-RV_{S\&L}$ of --1.1, 36.6, and --17.4\,\kms\ for neighbouring velocities of \citet{gar93}, from FEROS data of 2006, and for RVs of \citet{gom21}, respectively. These RVs were corrected for these shifts and also for the $\gamma=-9$\,\kms\ of the Stickland \& Lloyd solution. The Li\`ege orbital solution package (LOSP\footnote{https://www.stsci.edu/~hsana/losp.html}) was then used to derive the best orbital solution. It is provided in Table \ref{solorb} and shown on the right panel of Fig. \ref{rvphot}. The rms is 10\,\kms, and the agreement with the Stickland \& Lloyd solution is excellent. The mass function implies that for a primary of 33.9\,M$_{\odot}$, the typical mass of an O8.5I star \citep{mar05}, the secondary would have 5.4--5.8\,M$_{\odot}$ for inclinations of 90--70$^{\circ}$. This corresponds to mass ratios of $q=M_2/M_1=0.16-0.17$.

\citet{gom21} do not list $T_0$ (whether periastron or conjunction), $K$, nor $f(m)$, prohibiting comparison. They do provide eccentricity and periastron argument but the parameters strongly depend on the chosen model. In their Model 1, these authors fitted the photometry and spectroscopy together, leading to $e=0.28\pm0.01$ and $\omega=0.79\pm0.03$\,rad (i.e. $45.3\pm1.7^{\circ}$). In their Model 2, they fitted periastron argument separately for photometry and spectroscopy and derived $e=0.12\pm0.01$ with $\omega_{RV}=1.10\pm0.06$\,rad (i.e. $63.0\pm3.4^{\circ}$) and $\omega_{phot}=5.91\pm0.06$\,rad (i.e. $338.6\pm3.4^{\circ}$). The large eccentricity variation is not discussed but these authors attributed the large difference in the periastron argument to tidal effects. However, their photometric and spectroscopic datasets were taken in June 2015--April 2016 and April 2016--February 2017, respectively. The average difference between datasets thus is one year, but their individual duration is also of about one year. If apsidal motion was that large, namely, 1.5\,rad (or $85^{\circ}$) over a year, the spectroscopic and photometric datasets should not be combined; at the same time,   the photometric dataset nor the spectroscopic one should be analysed as a single entity either. Indeed, between the start and the end of each dataset, $\omega$ would have changed so much that the analysis should include the $\omega$ variations, which was not done. However, (1) apsidal motion in massive binaries usually appears much less extreme (examples of large values are 7.5$^{\circ}$\,yr$^{-1}$ in Y\,Cyg, \citep{har14}, and 15$^{\circ}$\,yr$^{-1}$ in CPD--41$^{\circ}$7742, \citealt{ros22}) and (2) high-quality light curves do not confirm the photometric interpretation of \citet{gom21}. More details are given in the next subsection.

Finally, a last note ought to be made on the FEROS spectra. While the relative strength of the  He\,{\sc i} and  He\,{\sc ii} lines confirms the O8.5 type, the spectra also display a Si\,{\sc iv}\,$\lambda$4088\AA\ line similar in intensity to the nearby He\,{\sc i}\,$\lambda$4143\AA. For late O-type stars, this suggests a main sequence classification, a luminosity class also adopted by \citet{gar93}. A more general comparison with the atlas of \citet{sot11} also favour a main sequence class. However, the absolute magnitude $M_V$ (derived in previous subsection) is 2\,mag too bright for a main sequence classification and the presence of faint companions (see \citealt{san14} and the next subsection) are not sufficient to explain such a large difference.

\begin{figure}
  \centering
  \includegraphics[width=9cm]{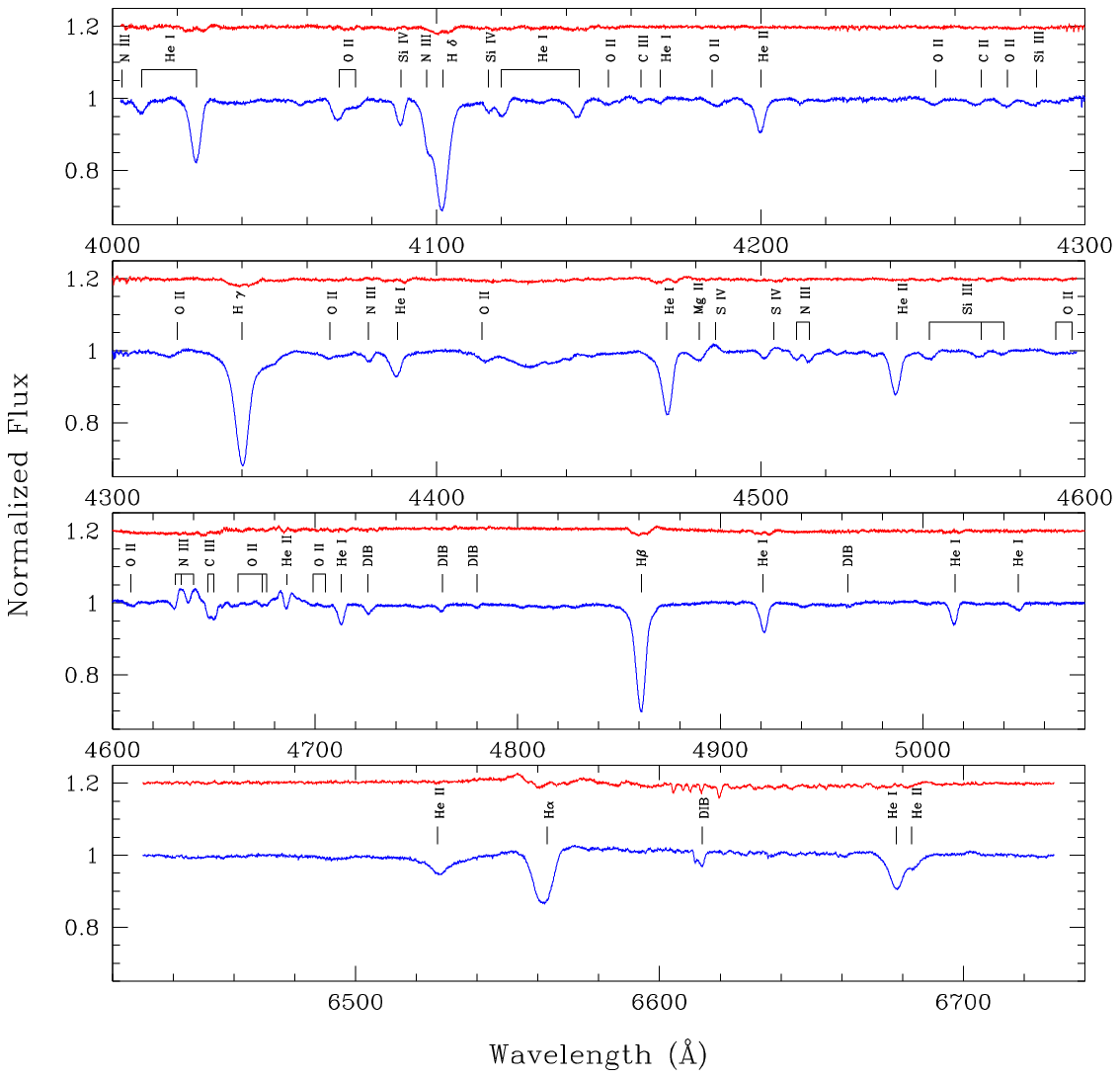}
  \caption{Results of the disentangling applied to the six FEROS spectra from 2006, keeping the RVs fixed and assuming $q = 0.18$. The reconstructed spectra of the primary (blue) and secondary (red) are normalized against the combined continuum of  the system. For clarity, the normalized secondary spectrum is shifted vertically by 0.2. 
  }
  \label{figdisent}
\end{figure}

To search for spectral signatures of the secondary star, we used our implementation of the shift-and-add spectral disentangling method described by \citet{gon06}. Spectral disentangling takes advantage of the Doppler shifts at different orbital phases to iteratively reconstruct the individual mean spectra of the components of a binary system. The \citet{gon06} method allows us (in principle) to simultaneously determine the individual RVs for each epoch of observation \citep[see e.g.][]{rau16}. However, in the present case, we fixed the RVs of the primary star to those determined for He\,{\sc ii} (see Table \ref{rv}). The RVs of the secondary were computed from our best estimate of the mass ratio $q = 0.18$ (see next subsection), considering the zero point of --27.6\,\kms\ mentioned above, and were also kept fixed in the disentangling process, as done in \citet{naz23}. This method was applied to data over the spectral ranges from 4000--5080\,\AA\ and 6430--6730\,\AA. We used 200 iterations, allowing to remove any residuals from the initial approximation of flat, featureless spectra. To avoid any shift in the systemic velocity that could arise on longer timescales because of the presence of a third component, we restricted ourselves to the six FEROS spectra that were taken within two months in 2006. The results are displayed in Fig. \ref{figdisent}.

We stress that the disentangled spectra, especially that of the primary, could suffer from contamination of the tertiary spectrum. We attempted to disentangle the spectra accounting for three components, but these attempts were not successful. Indeed, the RV amplitude of the primary is rather low, leading to severe cross-contamination of the primary and tertiary spectral components. We thus restricted ourselves to the disentangling of two spectral signatures (i.e. the primary and the secondary spectra). With those caveats in mind, we find that the reconstructed primary spectrum appears very much as we would expect for a late O-type star. Some remarkable features are the N\,{\sc iii}\,$\lambda$4634-4640 emissions, along with a double-peaked He\,{\sc ii}\,$\lambda$\,4686 emission. These features qualify the primary as an Of star and the double-peaked morphology of He\,{\sc ii}\,$\lambda$\,4686 even suggests an Onfp/Oef classification. However, the objects of the latter category are usually of earlier spectral types and are faster rotators \citep{rau21} than the primary of \hd. Indeed, we used the Fourier method to establish the projected rotational velocity $v\,\sin{i}$ of the primary \citep{sim07}. For this purpose, we used the C\,{\sc ii}\,$\lambda$\,4268, the O\,{\sc ii}\,$\lambda$\,4276 and the Si\,{\sc iii}\,$\lambda$\,4568 lines on the primary disentangled spectrum. We obtained a value of $v\,\sin(i) = (175 \pm 3)$\,km\,s$^{-1}$, which falls in between published values (135\,km\,s$^{-1}$ in \citealt{sti01} and $\sim$200\,km\,s$^{-1}$ in \citealt{gom21}).

\begin{figure*}
  \centering
  \includegraphics[width=9cm]{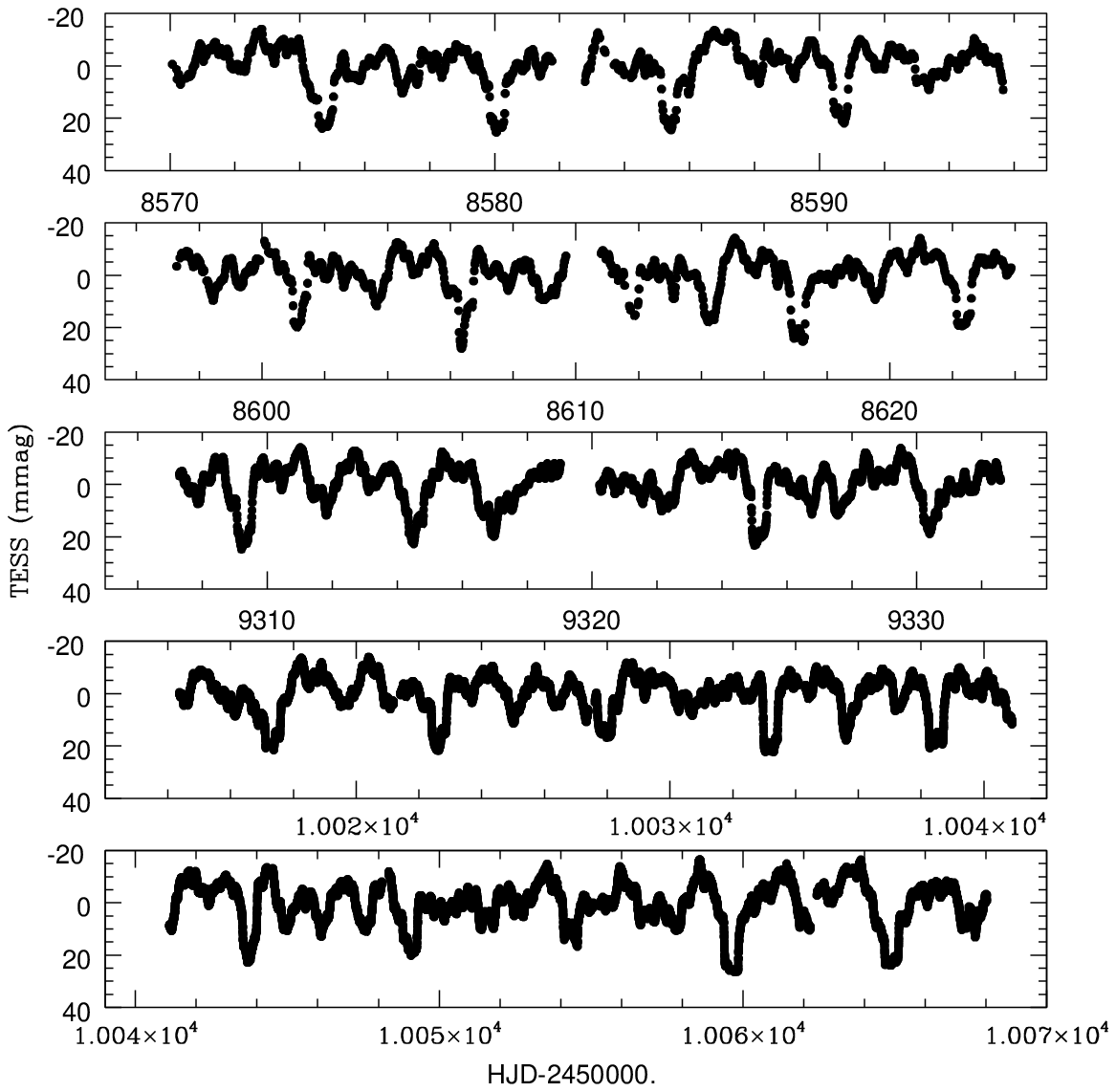}
  \includegraphics[width=9cm]{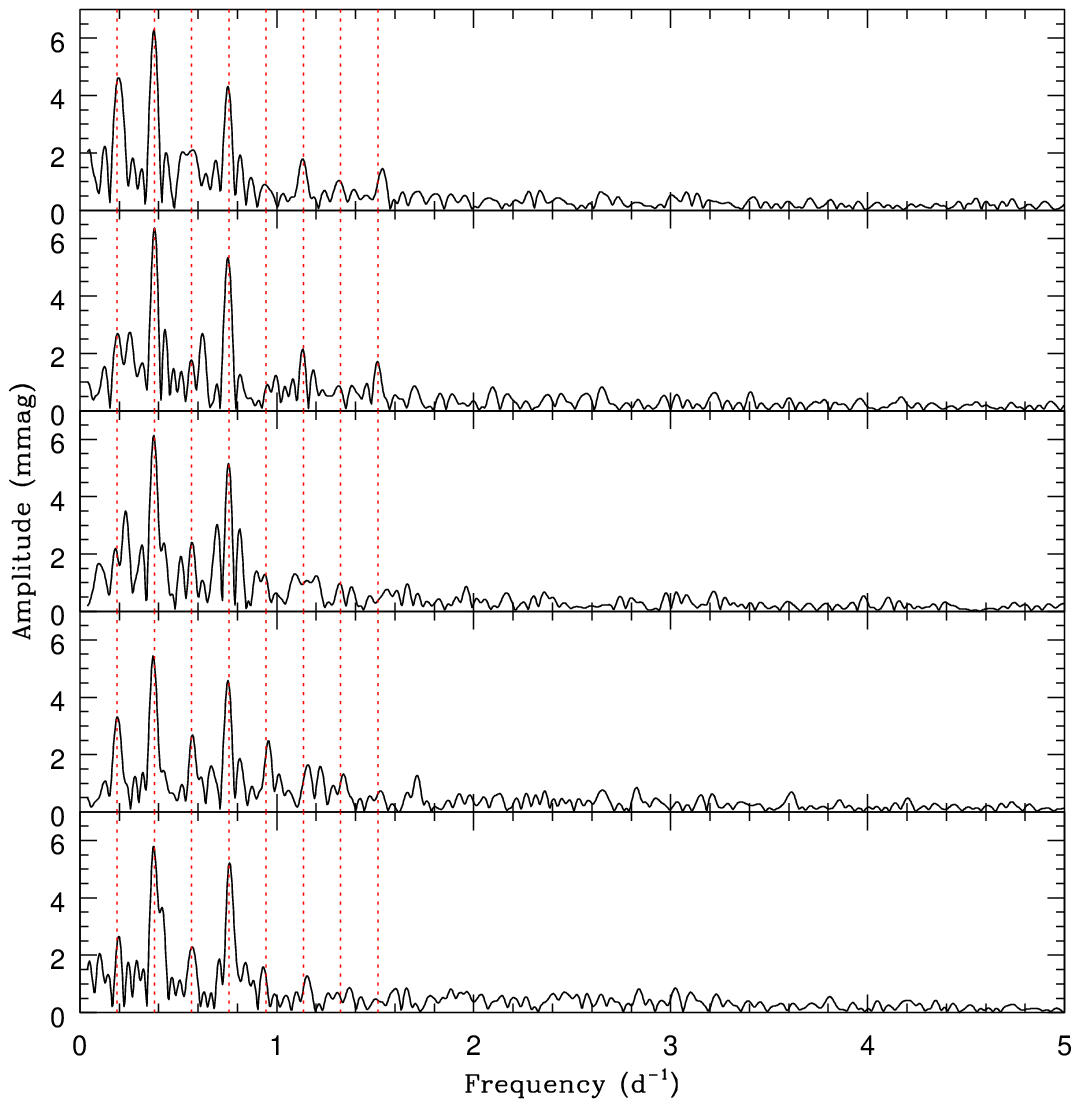}
  \caption{\te\ light curves of \hd, and their associated periodograms. Dotted red lines in the right panels indicate the orbital frequency and its harmonics.
  }
  \label{lc}
\end{figure*}

The reconstructed secondary spectra unveil only a limited number of lines. All H\,{\sc i} Balmer lines display relatively broad and shallow absorptions. Broad absorptions, possibly with an emission reversal in the core, are also seen in the strongest He\,{\sc i} lines (at 4026, 4471, 4921\AA), whereas, with the exception of He\,{\sc ii}\,$\lambda$\,4686, He\,{\sc ii} lines are absent from the secondary spectrum. H$\alpha$ exhibits a similar profile to the He\,{\sc i} lines possibly flanked by two broad emission features. Whilst these results must be considered as preliminary, we note that the spectral reconstructions are quite insensitive to $q$. Indeed, we tested mass-ratios between 0.16 and 0.20 and the reconstructed spectra were essentially identical.

We may, however, note some limitations of this disentangling trial. First, only six spectra could be used, which of course limits the quality of the result: disentangling should be re-tried with a much more intense monitoring (with also an increased S/N and redundancy). Second, there is an intrinsic limitation to the reconstruction. Indeed, the lines dominating the spectrum come from the primary spectrum and they are broader than its orbital motion (i.e. $v\,\sin(i)>K$). The contribution of the secondary and tertiary stars  most probably ends up partially confused with the primary contribution. This impairs the disentangling, which works best on well-separated signatures. 

Finally, we took a second look at the two IUE spectra\footnote{downloaded from the INES database, see http://sdc.cab.inta-csic.es/cgi-ines/IUEdbsMY} used by \citet{sti01}. After normalization, the IUE spectra were cross-correlated, considering wavelength ranges avoiding strong wind lines (e.g. C\,{\sc iv} near 1550\AA), against a TLUSTY OSTAR2002 spectrum \citep{lan03} with high temperature and high $\log(g)$, as in \citet{wan21}. While we could retrieve the peak in the cross-correlation function corresponding to the primary star, no specific feature was detected at the expected velocity of the secondary star. This non-detection is not totally surprising as the S/N of these IUE spectra was low and the temperature contrast between both components is lower than for Be+subdwarf systems, which makes detection more complex \citep{jon22}.  

\subsection{Constraints from the light curves}
The \te\ light curves are shown on the left side of Fig. \ref{lc}. Regular deeper dips of about 20\,mmag are readily detected, with a separation of about 5\,days (i.e. similar to the orbital period). There are, however, additional oscillations (notably, additional dips in between the main ones). The deeper dips and the intermediate ones clearly represent primary and secondary eclipses. In other words, this behaviour does not resemble the ellipsoidal variations (45\,mmag peak-to-peak) along with a flare of a similar amplitude described by \citet{gom21}. In fact, their ground-based data can be retrieved from the paper publisher website as well as at CDS\footnote{https://cdsarc.cds.unistra.fr/viz-bin/Cat?J/ApJ/913/48\#/browse}: it appears that they clearly suffered from a huge scatter, which explains the difficulty to interpret them (see  more details below). 

Frequency spectra were calculated for each Sector data using the same modified Fourier algorithm as in previous subsection \citep{HMM,gos01}. The frequency spectra clearly show peaks at the orbital frequency and its harmonics (see right panels of Fig. \ref{lc}). The strongest peak corresponds to the first harmonics at 0.376$\pm$0.004\,d$^{-1}$. There is no strong peak in addition to these harmonics. Therefore, no evidence for additional rotational modulation or pulsational activity has been found, but there clearly is red noise, as is common in massive stars \citep{bow20,naz21}.

\begin{figure*}
  \centering
  \includegraphics[width=6cm]{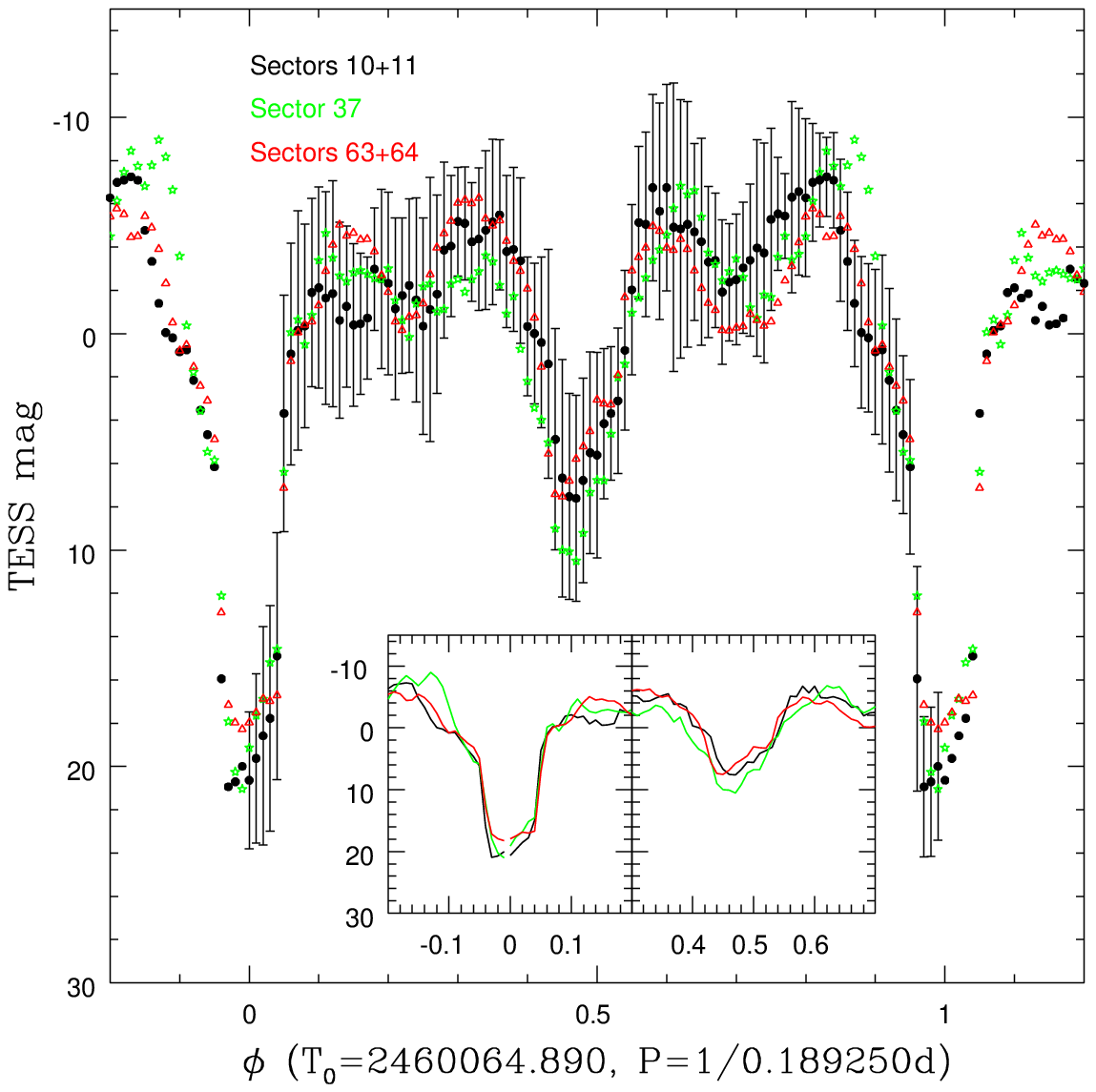}
  \includegraphics[width=6cm]{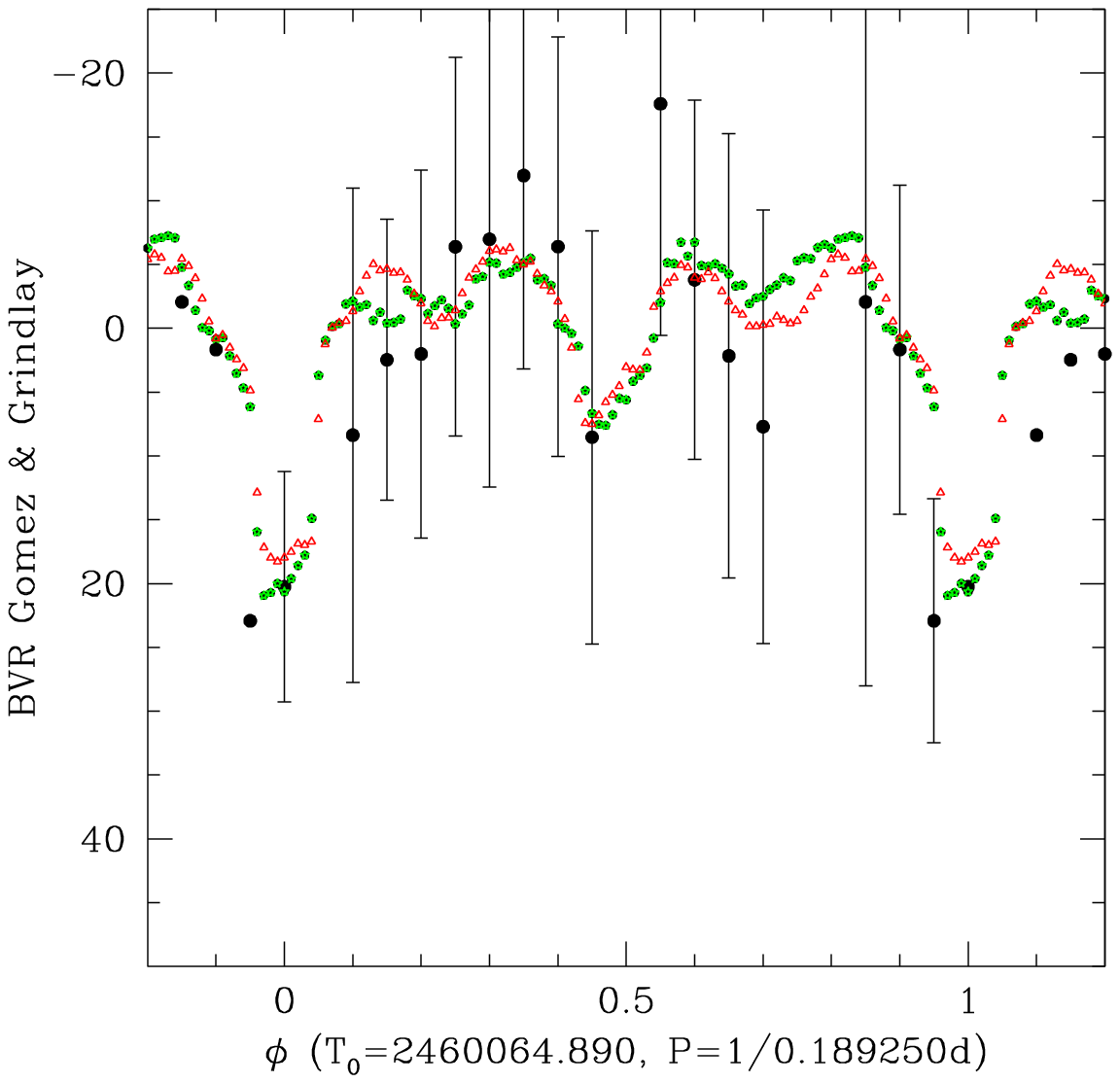}
  \includegraphics[width=6cm]{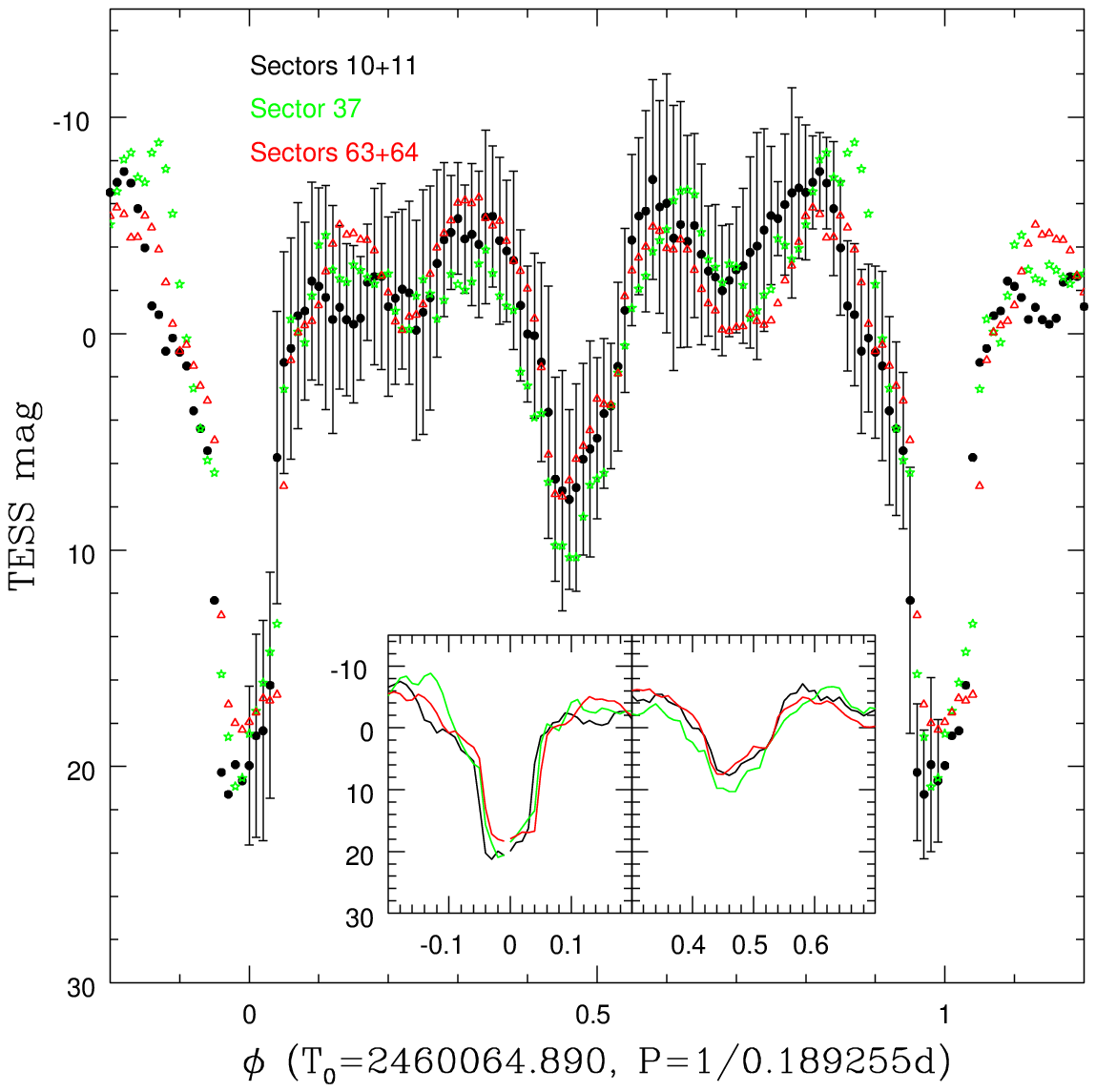}
  \caption{Binned light curves for the three \te\ observing campaigns (left) and for the Gomez \& Grindlay photometry (middle, a zero point of 7.62 was subtracted) using $f_{orb}=0.189250$\,d$^{-1}$. Error bars correspond to a 1$\sigma$ dispersion in each bin. Right: Same as left panel but using $f_{orb}=0.189255$\,d$^{-1}$ (the spectroscopic one).
  }
  \label{lc2}
\end{figure*}

The light curves of Sectors 10+11, 37, and 63+64 were folded using $T_0=2460064.890$ and an orbital frequency of 0.18925\,d$^{-1}$ which provides the best phasing of the deepest eclipses. This frequency is only slightly lower than that derived from spectroscopy ($0.189255$\,d$^{-1}$) and compatible with it, within the errors. These folded curves were then binned to get average phased light curves with 100 phase bins (left panel of Fig. \ref{lc2}). In the binned light curves, the stochastic `noise' is much lower; hence, the second eclipse is much more obvious. The main eclipses are well centered on $\phi=0,$ but the phasing may seem less good for the second eclipses (see insets of left panel of Fig. \ref{lc2}). When the orbital frequency derived from the spectroscopy (0.189255\,d$^{-1}$) is used instead, the results appear very similar, although the second eclipses agree well, while slight shifts are seen for main eclipses (see right panel of Fig. \ref{lc2} and its insets). These shifts are very small, $\Delta(\phi)=0.01$, but might be due to slow apsidal motion. However, they may also have a more mundane explanation. Indeed, the depth of the eclipses changes by about 2\,mmag in these average curves and additional dips also appear in between the eclipses. Furthermore, the typical dispersions around the average amount to $\sim5$\,mmag (Fig. \ref{lc2}). The orbital signal is thus nearly of the same order of magnitude as the additional variations (Fig. \ref{lc}), which may blur the eclipse profiles, giving an appearance of shifts. We therefore decided to use the spectroscopic orbital period for the light curve analysis.

We also binned the Gomez \& Grindlay $BVR$ data. Due to the coarser coverage of the orbital cycle, only 20 bins were used here. No distinction was made between the different filters as the filter information is missing in the available data. In addition, in \citet{gom21}, all $BVR$ data were considered together so the same approach was followed here. The middle panel of Fig. \ref{lc2} compares these binned curves to the average \te\ curves. Whatever the chosen orbital frequency, they appear fully compatible with each other. In particular, the eclipses are also present in the ground-based data, although the data display a huge dispersion: it is most likely that the low precision is what led to the previous erroneous interpretation. 

In Figure \ref{rvphot}, the folding is done using the photometric $T_0$ and the spectroscopic $P$. It may be noted that the main eclipse occurs when the RV of the primary star increases, namely, at the conjunction with the O star primary in front of its companion. Furthermore, the flat bottom of the eclipses suggest that the companion is fully eclipsed. Therefore, the fact that this deepest eclipse occurs when the O star occults its companion implies that this companion is hotter than the primary star. As we  show below, this conclusion is confirmed by fitting the light curves with a binary model.

To analyse the photometry, we built an overall mean light curve to help to get rid (as much as possible) of the stochastic variations and better isolate the orbital signal. To this aim, we did not consider the full set of data points. Indeed, because of the different cadences, the last two sectors have ten times more points than the first two sectors; hence, a simple average would actually be heavily biased towards the last observations. We therefore averaged datasets in each group of sectors (i.e. those of Sectors 10+11, 37, and 63+64) and then took the mean of these individual curves. We analysed the result with Nightfall v1.92 \citep{wic11}, assuming $I$ band for the data (the closest to \te\ bandpass) and a quadratic limb darkening. We took into account the presence of a third light ($\Delta_H=1.26$\,mag, \citealt{san14}, which gives $F_H^{3rd*}/F_H^{binary}=0.31$ or a luminosity ratio of $F_H^{3rd*}/F_H^{total}\sim0.24,$ which we applied to the I band). We also fixed the eccentricity to the value of 0.103 derived from spectroscopy and the primary temperature and mass to values typical for O8.5I stars (30.5\,kK and 33.9\,M$_{\odot}$, \citealt{mar05}).

\begin{figure}
  \centering
  \includegraphics[width=9cm]{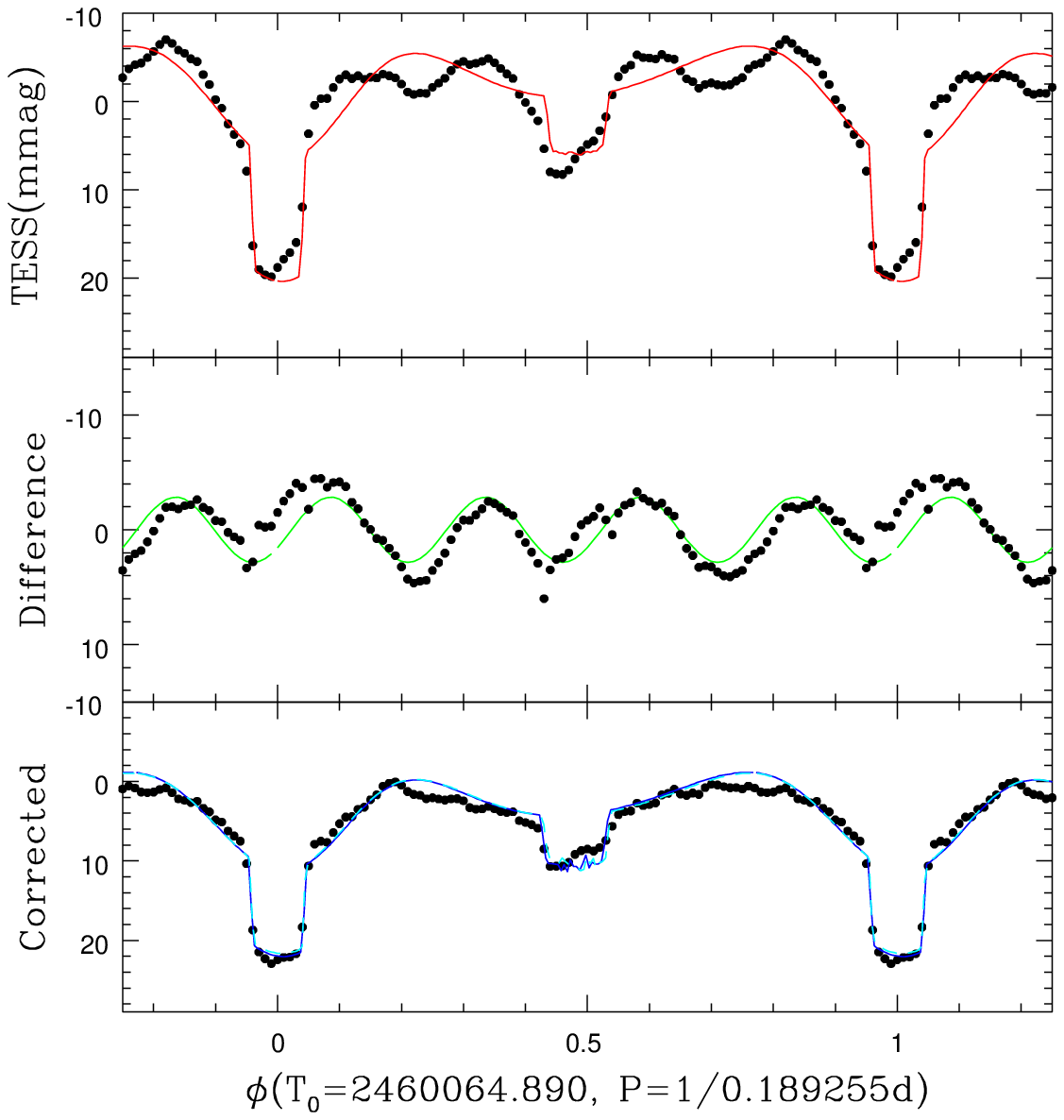}
  \caption{{\it Top:} Average light curve with the first best-fit photometric solution. {\it Middle:} Residuals and their best-fit sine wave. {\it Bottom:} Average light curve after removal of this sine wave, and best-fit photometric solution (solid blue line = supergiant solution, dashed cyan line = giant solution).
  }
  \label{phot}
\end{figure}

We performed several fitting trials using the range of mass ratios mentioned in the previous subsection. A good fit was found for a mass ratio $q=0.164,$ but the residuals clearly present a wavy appearance, which can already be spotted in the original light curve (Fig. \ref{phot}). They correspond to one of the harmonics detected in the frequency spectra, the unusually strong signal at $f=4\times f_{orb}$. These variations could reflect tidally excited oscillations which are not totally uncommon in binary systems, as, for instance, shown in \citet{kol21}. We decided to fit a sine wave with that harmonic frequency into the residuals, then we subtracted the best-fit sine wave from the data, and we repeated the two-step averaging procedure. 

The fitting was restarted on the cleaned light curve, using detailed reflection and performing a large exploration of the parameter space. The parameters of the best-fit photometric solutions are provided in the middle column of Table \ref{solorb} and shown in the bottom panel of Fig. \ref{phot}. The errors were calculated assuming the best-fit reduced $\chi^2$ corresponds to 1: the range of value for each parameter was then evaluated in all solutions corresponding to reduced $\chi^2=1+5.89/(N-5)$ (1-$\sigma$ for 5 fitted parameters) or $\chi^2=1+18.2/(N-5)$ (3-$\sigma$ range), where $N=100$, the number of bins in the average light curve. The best-fit periastron argument is in agreement (within the errors) with the spectroscopically derived value and the predicted amplitude of the primary RV curve is 57.1\,\kms, also in excellent agreement with the observed value. The primary radius is however nearly twice too small compared to usual radii of O8.5I supergiants \citep{mar05}. It better agrees with a giant classification. The primary bolometric luminosity, derived from its temperature and radius, amounts to $\log(L/L_{\odot})=5.12$: it is close to that expected for giant stars, but nearly three times lower than for supergiants \citep{mar05}. 

Therefore, if we instead consider typical parameters of giant O8.5 stars (temperature of 31.7\,kK and mass of 24.84\,M$_{\odot}$, \citealt{mar05}), the mass function from the spectroscopic solution implies secondary masses of 4.5--4.8\,M$_{\odot}$ for inclinations of 90--70$^{\circ}$ (i.e. $q=M_2/M_1=0.18-0.19$). The fitting was thus restarted considering these values and the best-fit parameters are provided in the last column of Table \ref{solorb}. The predicted amplitude of the primary RV curve now is 57.5\,\kms, and the bolometric luminosities are $\log(L/L_{\odot})=5.07$ and 3.71 for the primary and secondary, respectively. 

In both cases, if we add the primary luminosity, the companion luminosity (about 4\% of the primary luminosity), and the contribution of the third light (about a quarter of the total luminosity, see above), we get a total luminosity of $\log(L/L_{\odot})\sim5.2$. This luminosity still is $\sim$50\% of that derived from the $V$ magnitude, using the bolometric correction of O8.5 giant and supergiant stars and the smallest extinction and distance values.

Finally, we performed a fitting trial fixing the secondary temperature to 15\,kK, typical of B5--6V stars with  masses similar to that derived for the companion. While the fitting converges towards similar filling factors, inclination, and periastron argument, the deepest eclipse cannot be reproduced hence such a fitting remains unacceptable. 

\begin{table}
  \scriptsize
  \caption{Orbital solution for \hd.
 \label{solorb}}
  \begin{tabular}{lll}
    \hline
\multicolumn{3}{l}{\it Spectroscopic solution}\\
    Parameter & This work & \citet{sti01}\\
\hline
$\gamma$ & $-0.4\pm2.1$\,\kms\ & $-9.0\pm1.7$\,\kms \\
$K$      & $57.6\pm2.8$\,\kms\ & $56.8\pm1.9$\,\kms \\
$e$      & $0.103\pm0.046$     & $0.069\pm0.037$ \\
$\omega$ & $278\pm33^{\circ}$   & $266\pm34^{\circ}$ \\
$f(m)$   & $0.103\pm0.015$\,M$_{\odot}$    & $0.104\pm0.011$\,M$_{\odot}$ \\
$P$      & $5.28388\pm0.00018$d & $5.52963\pm0.00018$d \\
$T_0$(peri)&$2\,460\,065.006\pm0.483$   & $2\,446\,151.93\pm0.51$ \\
\hline
\multicolumn{3}{l}{\it Photometric solution}\\
Parameter & Supergiant & Giant\\
\hline
$T_0$(conj)&\multicolumn{2}{c}{2\,460\,064.890}   \\
$e$ & \multicolumn{2}{c}{0.103 (fixed)} \\
$q$ & 0.159 (fixed) & 0.179 (fixed) \\
$i$ & 84.6$\pm$3.0$^{\circ}$ & 87.7$\pm$1.9$^{\circ}$ \\
$\omega$ & 288$\pm$2$^{\circ}$ & 285$\pm$3$^{\circ}$ \\
$M_p$ & 33.90\,M$_{\odot}$ (fixed) & 24.84\,M$_{\odot}$ (fixed)\\
$T_{eff}^p$ & 30.5\,kK (fixed) & 31.7\,kK (fixed) \\
$R_p$ & 12.9$\pm$0.1\,R$_{\odot}$ & 11.3$\pm$0.1\,R$_{\odot}$ \\
$f_p$ & 0.659$\pm$0.007 & 0.645$\pm$0.008 \\
$M_s$ & 5.39\,M$_{\odot}$ (fixed) & 4.45\,M$_{\odot}$ (fixed)\\
$T_{eff}^s$ & 45.8$\pm$5.0\,kK & 49.9$\pm$4.0\,kK \\
$R_s$ & 1.14$\pm$0.03\,R$_{\odot}$ & 0.96$\pm$0.03\,R$_{\odot}$ \\
$f_s$ & 0.135$\pm$0.003 & 0.120$\pm$0.004 \\
\hline
  \end{tabular}
%\tablefoot{ }
\end{table}

\subsection{Constraints from the X-ray analysis}
\hd\ was detected as an X-ray source (XMMSL2 J110714.5--595226) in the XMM slew survey with just four counts, yielding an EPIC-pn count rate of $0.82\pm0.41$\,cts\,s$^{-1}$ in the 0.2--12.\,keV energy band (i.e. it is a 2$\sigma$ detection). It was subsequently re-observed by {\it NuSTAR}. Three exposures were taken (ObsID=30001050002, 4, and 6), with durations between 1 and 2 days, that is, sampling a large part of the orbit. The dates at mid-exposures ($HJD=2\,457\,086.469$, 111.663, and 123.226) correspond to phases of 0.32, 0.09, and 0.28, using our adopted ephemeris ($f_{orb}=0.189255$\,d$^{-1}$ and $T_0(conj)=2460064.890$). \citet{gom21} reported the fitting of the combined data with a thermal model ($kT=5.2$\,keV) as well as with a power law ($\Gamma=2.6$), deriving luminosities of $\sim 2\times10^{32}$\,erg\,s$^{-1}$ in the 2.--10.\,keV range. It is not known whether this value corresponds to the observed luminosity or has been corrected for any intervening absorption, but absorption correction is always small for such energies. Considering their distance (2.83\,kpc), this corresponds to an (average) flux of $\sim 2\times10^{-13}$\,erg\,cm$^{-2}$\,s$^{-1}$. 

{\it Chandra} observed \hd\ in October 2020 ($HJD=2\,459\,146.154$, or $\phi=0.12$). The reddening, $E(B-V),$ of 0.39--0.49\,mag corresponds to an absorbing column of 0.24--0.30$\times 10^{22}$\,cm$^{-2}$ using the formula of \citet{gud12}. A thermal model ($phabs_{ISM}\times phabs \times \sum apec$) requires two temperatures, but no additional absorption, to achieve a good fit: $N_{H, ISM}=0.24\times 10^{22}$\,cm$^{-2}$, $N_{H}=0$ (fixed), $kT_1=0.70\pm0.17$\,keV, $norm_1=(1.97\pm0.62)\times 10^{-4}$\,cm$^{-5}$, $kT_2=2.88\pm0.56$\,keV, $norm_2=(2.95\pm0.37)\times 10^{-4}$\,cm$^{-5}$, $\chi^2=38.51$ for 45 degrees of freedom. The observed fluxes are $(5.87\pm0.46)$ and $(2.04\pm0.26)\times10^{-13}$\,erg\,cm$^{-2}$\,s$^{-1}$ in the 0.5--10.\,keV and 2.--10.\,keV bands, respectively. For a power law fitting ($phabs_{ISM}\times phabs \times pow$), we would instead get: $N_{H, ISM}=0.24\times 10^{22}$\,cm$^{-2}$, $N_{H}=0$ (fixed), $\Gamma=2.98\pm0.14$, $norm=(2.82\pm0.24)\times 10^{-4}$\,cm$^{-5}$, and $\chi^2=61.89$ for 49 degrees of freedom. Observed fluxes are $(5.43\pm0.24)$ and $(1.80\pm0.20)\times10^{-13}$\,erg\,cm$^{-2}$\,s$^{-1}$ in 0.5--10.\,keV and 2.--10.\,keV, respectively. Those values agree well with the  {\it NuSTAR} results obtained from the hard band only. It is difficult to distinguish between the thermal and non-thermal scenarios using the available data. Indeed, the iron lines near 6.7\,keV (a tell-tale sign of the presence of hot plasma) can only be spotted if there are enough counts at high energies. Unfortunately, this is  not the case here.

The thermal fitting provides a better $\chi^2$ and its two temperatures appear rather similar to those found for `normal' OB stars, although the hot component here is very strong. In addition, the fluxes corrected for interstellar absorption amount to $8.4-8.6\times 10^{-13}$ \,erg\,cm$^{-2}$\,s$^{-1}$ in the 0.5--10.\,keV band, depending on the chosen model. With the bolometric luminosity of $\log(L/L_{\odot})=5.07$ and the smallest distance, this would yield a $\log(L_X/L_{BOL})$ of --5.8, about one dex above the usual value of --7 for `normal' massive stars, but much lower than recorded for high-mass X-ray binaries (in outburst). Certainly, the X-rays cannot be due to the sole intrinsic embedded wind-shocks of the primary O star.

\begin{figure}
  \centering
  \includegraphics[width=9cm]{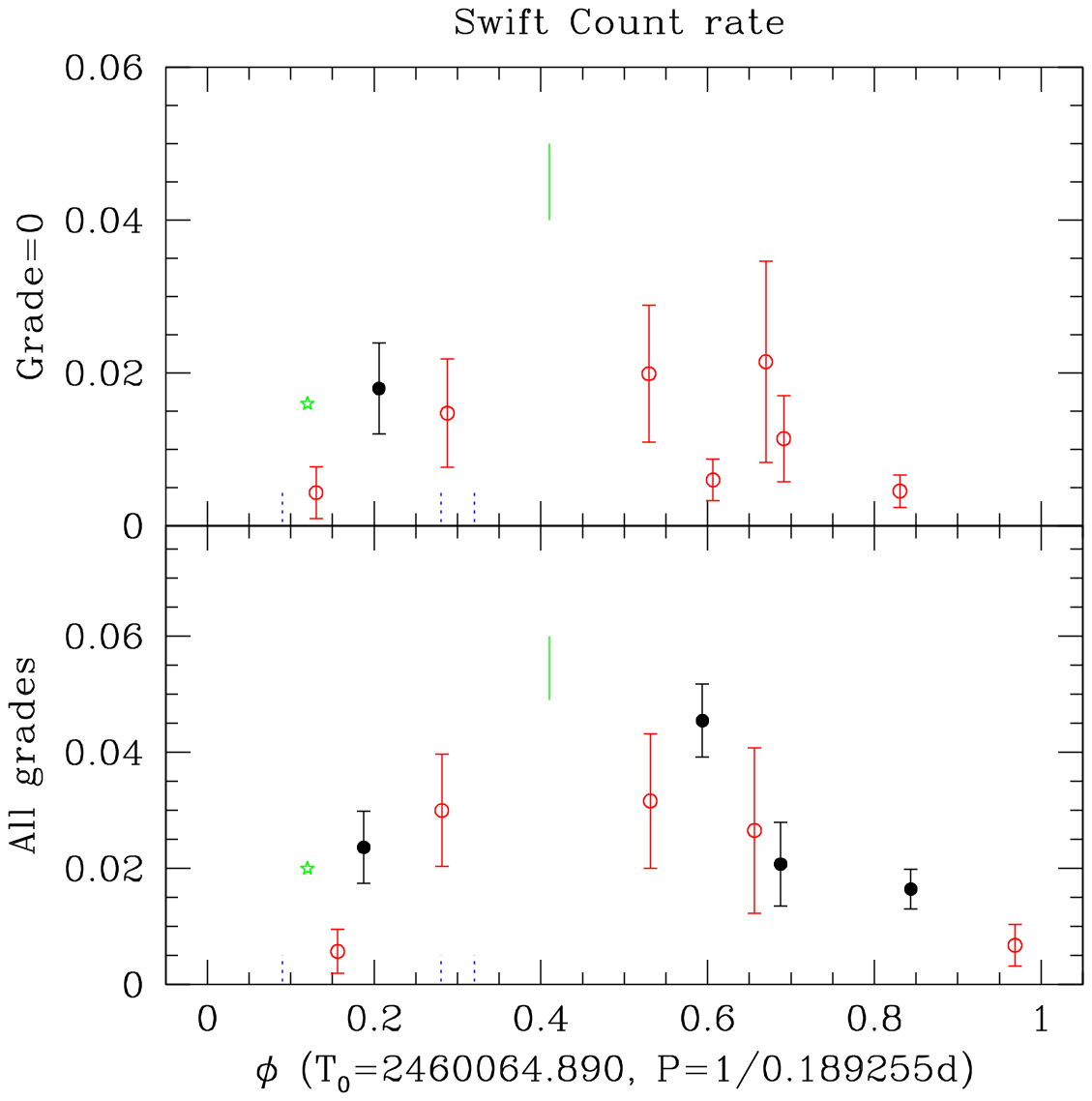}
  \caption{Evolution with orbital phase of the {\it Swift} count rates in the 0.5--10.\,keV band, with the more uncertain values in red (top: for $grade=0$ only, bottom: for all grades). Blue tick marks at the bottom indicate the phases of the {\it NuSTAR} observations. The green star and the green line indicate the {\it Chandra} and XMM slew points, respectively, converted to {\it Swift} values.
  }
  \label{swfig}
\end{figure}

\begin{table}
  \scriptsize
  \caption{{\it Swift} count rates in the 0.5--10.\,keV band.
 \label{sw}}
  \begin{tabular}{llll}
    \hline
    ObsID & $HJD$ & \multicolumn{2}{c}{Count rate ($10^{-2}$\,cts\,s$^{-1}$)}\\
          &       & $grade=0$ & all grades \\
\hline
00081443001 &2\,457\,110.309 & 0.46$\pm$0.21: &1.64$\pm$0.34  \\
00081443002 &2\,457\,111.174 & $<0.65$:       &0.68$\pm$0.36: \\
00081443003 &2\,457\,122.466 & 0.44$\pm$0.34: &5.70$\pm$0.38: \\
00081443004 &2\,457\,122.859 & 1.80$\pm$0.60  &2.36$\pm$0.62  \\
00015878001 &2\,459\,994.120 & 0.60$\pm$0.27: &4.55$\pm$0.63  \\
00087839003 &2\,460\,327.338 & 2.15$\pm$1.32: &2.65$\pm$1.43: \\
03103083001 &2\,460\,331.884 & 1.99$\pm$0.89: &3.16$\pm$1.16: \\
00087839004 &2\,460\,378.158 & 1.48$\pm$0.71: &3.00$\pm$0.97: \\
03103083004 &2\,460\,570.510 & 1.14$\pm$0.56: &2.07$\pm$0.72  \\
\hline
  \end{tabular}
\tablefoot{As exposures were short and the source was faint, an especially low number of count was collected in some exposures, which prohibits the refined centroiding and renders the count rates more uncertain: the associated data are maked by `:' in this table and are shown in red in Fig. \ref{swfig}.}
\end{table}

The {\it Swift} data of \hd\ were taken at several epochs. It was notably the main target of the exposures taken around the time of the {\it NuSTAR} observations. The most secure data correspond to $grade=0,$ but this additional filtering further reduces the number of counts: there are only 50 counts in the {\it Swift} spectrum, making the determination of spectral parameters quite uncertain. A trial was nevertheless attempted. A fitting using a single thermal model yields: $N_{H, ISM}=0.24\times 10^{22}$\,cm$^{-2}$ and $N_{H}=0$ (both fixed), $kT=0.82\pm0.13$\,keV, $norm=(1.10\pm0.18)\times 10^{-4}$\,cm$^{-5}$, $\chi^2=5.13$ for 3 degrees of freedom, observed fluxes of $(1.50\pm0.27),$ and $(0.12\pm0.04)\times10^{-13}$\,erg\,cm$^{-2}$\,s$^{-1}$ in the 0.5--10.\,keV and 2.--10.\,keV bands, respectively. A power law fitting yields: $N_{H, ISM}=0.24\times 10^{22}$\,cm$^{-2}$, $N_{H}=0$ (fixed), $\Gamma=2.62\pm0.38$, $norm=(1.03\pm0.17)\times 10^{-4}$\,cm$^{-5}$, $\chi^2=6.59$ for 3 degrees of freedom, observed fluxes of $(2.42\pm0.90),$ and $(1.07\pm0.63)\times10^{-13}$\,erg\,cm$^{-2}$\,s$^{-1}$ in 0.5--10.\,keV and 2.--10.\,keV, respectively. This latter fitting formally is less good than the thermal fitting, but it allows us to catch flux at higher energies (the thermal model fits only the coolest component detected by {\it Chandra}, as the sensitivity severely drops at high energies).

While uncertain, it may be noted that those parameters are of the same order of magnitude as those found from {\it Chandra} data. As these are broad-band measurements, the count rates should provide more secure information than spectra, at the expense of the spectral resolution. The individual {\it Swift} count rates are provided in Table \ref{sw} and shown in Fig. \ref{swfig}. Despite uncertainties, \hd\ seems to display a coherent, phase-locked behaviour, with a larger X-ray flux around $\phi=0.5$ than around $\phi=0$. This conclusion is backed up by the comparison with the XMM slew survey and {\it Chandra} results. Indeed, the best-fit power-law fitting of the {\it Chandra} spectrum converts\footnote{https://heasarc.gsfc.nasa.gov/cgi-bin/Tools/w3pimms/w3pimms.pl} to an EPIC-pn count rate of $\sim$0.3\,cts\,s$^{-1}$ in the 0.2--12.\,keV band, while the slew catalog provides a value more than twice larger. Considering our ephemeris, the {\it Chandra} observations were taken at a phase of 0.12, while the slew exposure is dated from 23 July 2010 ($HJD=2\,455\,401.392$); hence, it has a phase of 0.41. Figure \ref{swfig} further demonstrates this difference graphically as it displays the equivalent {\it Swift} count rates of both the {\it Chandra} flux and the XMM slew count rate (using the same {\it Chandra} model). All of these results must certainly be confirmed with higher-quality follow-up observations, but it provides a good hint that the X-ray flux of \hd\ might be variable by a factor of a few over its orbit.

\begin{figure}
  \centering
  \includegraphics[width=9cm]{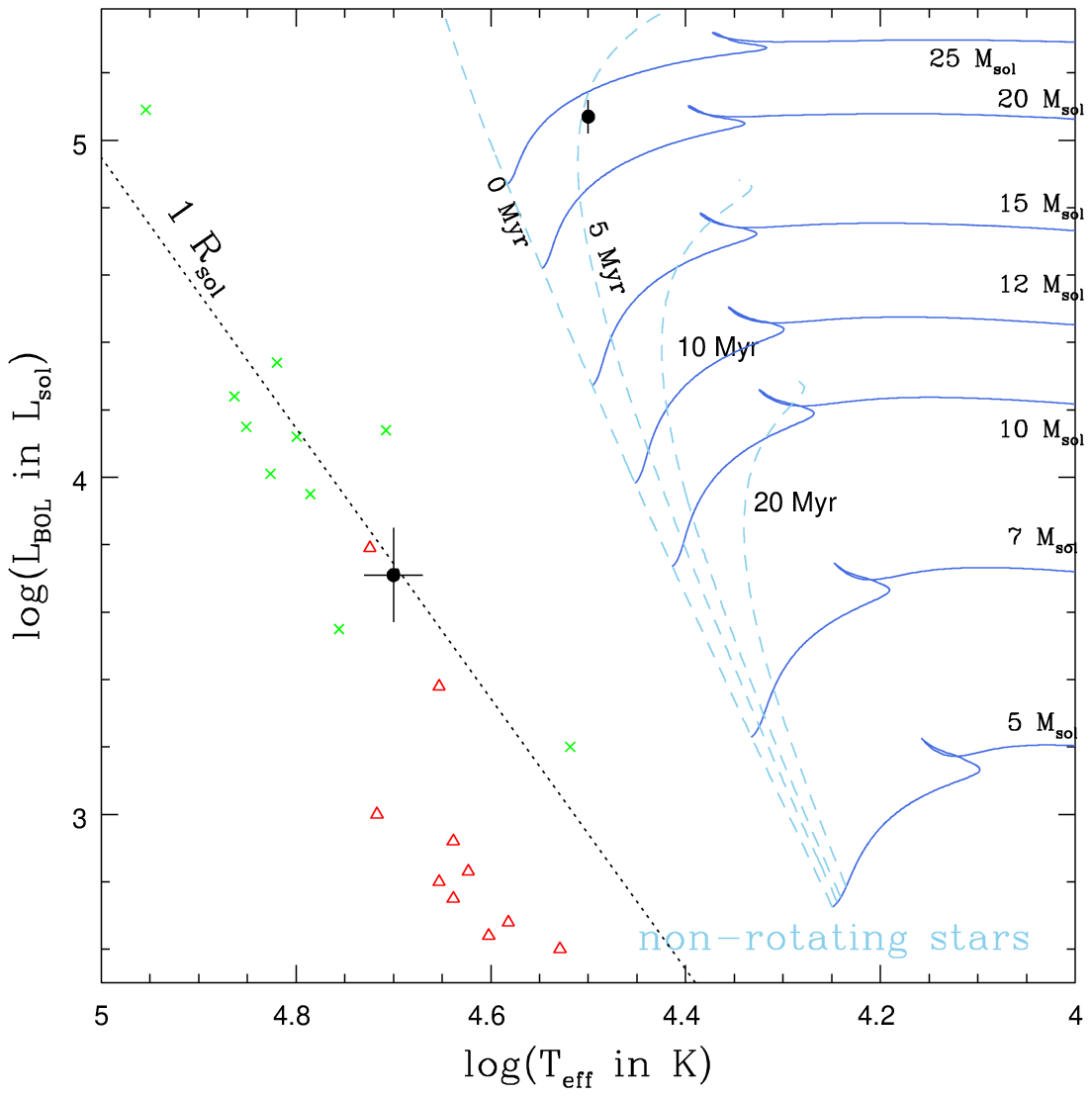}
  \caption{HRD showing the position of the two stars (black dots, Table \ref{solorb}) along with evolutionary tracks of \citet{bro11} for the Milky Way and no rotation. Stripped stars properties from \citet[and references therein]{naz22} are shown with red triangles and those from \citet{goe23} with green crosses.
  }
  \label{hrd}
\end{figure}

\section{Discussion}
It is evident that \hd\ harbours no black hole, as its light curve displays two eclipses. Instead, the companion is found to be a small, intermediate-mass, and hot object. While its mass may be compatible with a mid B-star, the high temperature and small radius cannot be reconciled with such an object. Such characteristics are much more typical of post-binary interaction systems. For example, the radius and temperature are similar to those observed in known (fully) stripped stars \citep{wan21}. We refer, in particular, to the the case of $\phi$\,Per, a system containing a Be star and a sdO companion (although the latter object has a smaller mass and a longer period than the secondary of HD\,96670 - $R_2=0.9$\,R$_{\odot}$, $T_2=53$\,kK, $M_2=1.2$\,M$_{\odot}$, $P=127$\,d, \citealt{mou15}). 

\begin{figure}
  \centering
  \includegraphics[width=9cm]{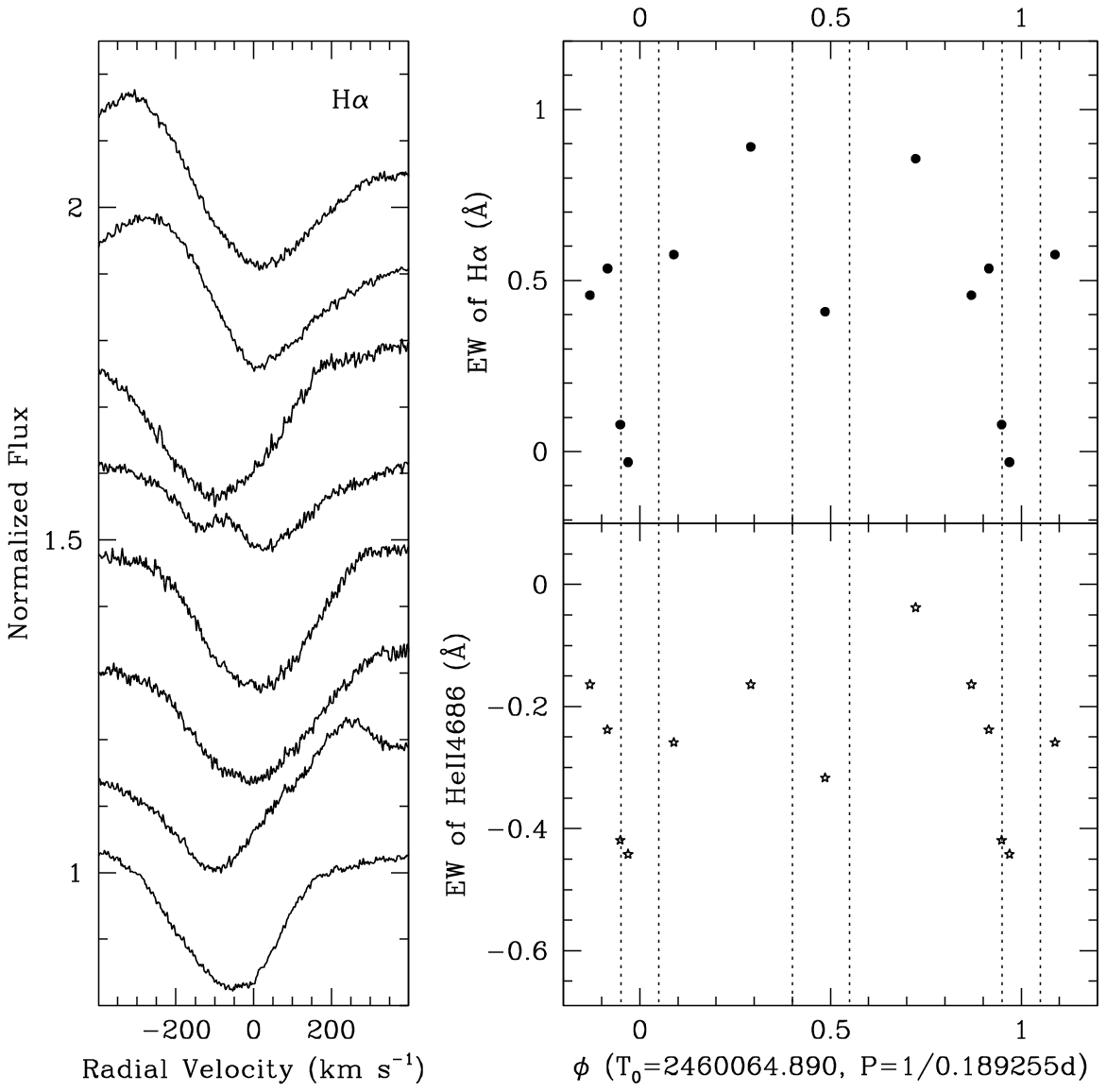}
  \caption{{\it Left:} Same as Fig. \ref{raies} for H$\alpha$. Time runs downwards (see Table \ref{rv} for exact dates); associated phases are, from top to bottom, 0.97 (2004 data), 0.95, 0.72, 0.92, 0.09, 0.29, 0.49 (2006 data), 0.87 (2009 data). {\it Right:} Equivalent widths of H$\alpha$ (top, black dots) and He\,{\sc ii}\,$\lambda$4686\AA\ (bottom, stars) as a function of phase. These EWs correspond to integration of the profiles in 6545.0--6575.0\AA\ and 4676.0--4693.5\AA, respectively. The vertical dotted lines correspond to the eclipse limits.
  }
  \label{specew}
\end{figure}

Figure \ref{hrd} shows the position of the two components of HD\,96670 in the Hertzsprung-Russell diagram (HRD). The position of the primary star in the HRD is compatible with models for slightly evolved 25\,M$_{\odot}$ objects. The primary appears close to the isochrone at 5\,Myr, in line with the age of Car\,OB2 from \citet{gar94}. In contrast, the secondary star is located well below the zero age main-sequence hence it is incompatible with single-star evolutionary tracks. However, it appears within the group of known fully stripped stars (see in Fig. \ref{hrd})\footnote{The intermediate evolutionary stages such as, e.g., the partially stripped cases reported in \citealt{ram24} (see their Fig. 16) are not shown here.}. Two different populations form this group in this figure. On the one hand, there are Galactic Be+sdOB systems for which the stripped star is much fainter than its Be companion and has $\sim$1\,M$_{\odot}$ and $T_{\rm eff}=17-53$\,kK. Their properties were (mostly) constrained through UV spectroscopy,  but also via interferometric campaigns and optical spectroscopy (\citealt{mou15,wan21}; see a summary in \citealt{naz22}, and references therein). Orbital solutions have been derived for about half of them. On the other hand, there are LMC systems where the stripped star dominates the output flux and has 1--8\,M$_{\odot}$ and $T_{\rm eff}=33-90$\,kK. These stripped stars were detected thanks to a UV photometric excess and their properties were derived from fitting optical spectra with atmosphere models \citep{goe23}. There is no UV spectroscopy or orbital solution available yet for these systems. Considering its mass and temperature, \hd\ appears to be an intermediate case between these two groups.

Turning to binary evolutionary models, \citet{goe18} presented a range of stripped star cases. The position of our secondary in the HRD is compatible with those found by these authors, for instance, in the core helium-burning phase of a stripped star with an initial mass of 9\,M$_{\odot}$. This specific model ends up with a stripped star of lower mass, however, \citet{goe18} presented only a limited set of cases (e.g. single mass ratio and single orbital period). A more extensive set can be found in the Binary Population and Spectral Synthesis (BPASS) v2.2.1 database \citep{eld17,ste20}. This database was searched for systems with properties similar to ours: solar metallicity ($Z=0.02$), period between 3 and 7\,d, primary mass between 3 and 7\,M$_{\odot}$, secondary mass between 20 and 30\,M$_{\odot}$, and primary radius between 0.3 and 3\,R$_{\odot}$. We note that the definition of primary and secondary in BPASS is inverted compared to ours since `their' primary corresponds to the initially massive star which has now become the least massive, post-interacting object (i.e. `our' secondary). One BPASS model fulfilled the criteria, namely, a system that started with a 17+12\,M$_{\odot}$ pair in a 2.7\,d orbit. After 11\,Myr, the period becomes 6.9\,d. The mass donor now is a stripped star with a temperature of 56--71\,kK, a radius of 1.7--2.9\,R$_{\odot}$, and a mass of $\sim$6\,M$_{\odot}$ while its companion has a temperature of $\sim$22.4\,kK, a radius of $\sim$8.2\,R$_{\odot}$, and a mass of $\sim$20.8\,M$_{\odot}$. Of course, this is not a perfect match to the \hd\ system, but the database remains generic; thus, already finding a relatively close case is a very encouraging start. In this context, we may note that the secondary of \hd\ also falls amongst the stripped helium stars of the simulations by \citet{yun24}, and in particular their Fig. 4. Although it is beyond the scope of this paper, a more detailed modelling that is specifically tuned to the case of \hd \ will certainly lead to a better match.

\citet{goe18} calculated typical spectra for stripped stars with a range of temperatures and masses. For 4--5\,M$_{\odot}$ cases, the spectra display a strong and broad emission in He\,{\sc ii}\,$\lambda$\,4686\AA, small emission components in Balmer lines, and no line at He\,{\sc ii}\,$\lambda$\,4542\AA. While the secondary of HD\,96670 appears cooler than Goetberg's stars, this may explain why the He\,{\sc ii}\,$\lambda$\,4542\AA\ remains uncontaminated, while the profile of He\,{\sc ii}\,$\lambda$\,4686\AA\ is complex. Clearly, a denser monitoring of the orbital cycle is needed to understand the behaviour of He\,{\sc ii}\,$\lambda$\,4686\AA.

While the nature of the system now seems quite clear, a question remains regarding the origin of the X-ray emission. Both massive OB stars and hot subdwarf stars are high-energy sources, but their X-rays are soft and faint, following $\log(L_X/L_{BOL})\sim -7$ \citep{naz11,mer16}. This is clearly not the case of HD\,96670. Regarding stripped stars paired with massive stars, only Be+sdO systems have been examined up to now and no particularly hard or bright X-rays were detected for them\footnote{Except for the subgroup of $\gamma$\,Cas analogs, which do not display phase-locked changes.} \citep{naz22}. However, O stars have much stronger winds than the Be stars examined in that paper, while the stripped companion here is  more massive. The bright and hard character of the X-rays, coupled to the phase-locked variations hinted by current data, are reminiscent of colliding wind phenomena (for a review, see \citealt{rau22}). In particular, the X-ray bright part of the colliding winds may be occulted by a large stellar body in high-inclination systems: this could explain the flux decrease seen at $\phi=0$ in \hd. Of course, whether and exactly how a colliding wind phenomenon applies to \hd\ remains to be demonstrated with a sensitive monitoring over the orbital cycle. In addition, a sensitive X-ray exposure would be able to clarify the nature of the high-energy emission; in particular, the presence of a non-thermal component versus a hot thermal component, which would also strongly constrain its origin.

The hypothesis of a wind-wind collision is reinforced by the analysis of the equivalent widths (EW) of the two lines presenting emission components, H$\alpha$ and He\,{\sc ii}\,$\lambda$4686\AA. While the equivalent widths of He\,{\sc i} and other He\,{\sc ii} lines appear rather stable with phase in the FEROS spectra, both contaminated lines show a decrease in their EWs during eclipses (right panel of Fig. \ref{specew}; the effect appears smaller in H$\beta$). After an examination of the line profiles (left panels of Figures \ref{raies} and \ref{specew}), it appears that this EW change clearly comes from an increased emission, not from a decreased absorption. This can be understood in the context of colliding winds. Indeed, the shocked plasma is not at the same temperature everywhere in the collision zone, with the highest temperatures (typical of X-ray emitting plasma) usually reached in a small zone near the line-of-centers, while optical lines arise in a very large zone further away from the stars. Such a configuration can explain both a smaller X-ray emission when the large O star hides the small region with the hottest plasma and larger emission components in the optical lines (as the associated emission is diluted by a smaller continuum). These conclusions are encouraging for the colliding wind model, but they remain preliminary, owing to the small number of spectra and the sparse phase coverage. 

In conclusion, more studies on HD\,96670 are required, but at least the basic parameters of the system have been established. Clearly, additional optical and X-ray monitoring should be performed to improve our understanding of this system and better pinpoint the past and/or current interactions between its components. 

\begin{acknowledgements}
The authors thank the referee for helpful comments, the Swift helpdesk for their help, and Queen's music for inspiration (hence the title as hommage). They also acknowledge support from the Fonds National de la Recherche Scientifique (Belgium), the European Space Agency (ESA) and the Belgian Federal Science Policy Office (BELSPO) in the framework of the PRODEX Programme (contracts linked to XMM-Newton and Gaia). This paper includes data collected by the TESS mission, which are publicly available from the Mikulski Archive for Space Telescopes (MAST). Funding for the TESS mission is provided by NASA's Science Mission directorate. ADS and CDS were used for preparing this document. 
\end{acknowledgements}

\end{document}